\documentclass[sigconf]{acmart}

\AtBeginDocument{%
  \providecommand\BibTeX{{%
    \normalfont B\kern-0.5em{\scshape i\kern-0.25em b}\kern-0.8em\TeX}}}

\copyrightyear{2021}
\acmYear{2021}
\setcopyright{acmlicensed}\acmConference[CHI '21]{CHI Conference on Human Factors in Computing Systems}{May 8--13, 2021}{Yokohama, Japan}
\acmBooktitle{CHI Conference on Human Factors in Computing Systems (CHI '21), May 8--13, 2021, Yokohama, Japan}
\acmPrice{15.00}
\acmDOI{10.1145/3411764.3445779}
\acmISBN{978-1-4503-8096-6/21/05}

\usepackage{subcaption}

\begin{document}

\title{Dark Patterns and the Legal Requirements of Consent Banners: An Interaction Criticism Perspective}


\author{Colin M. Gray}
\email{gray42@purdue.edu}
\orcid{0002-7307-1550}
\affiliation{%
  \institution{Purdue University}
  \streetaddress{401 N Grant Street}
  \city{West Lafayette}
  \state{Indiana}
  \postcode{47907}
  \country{United States}
}

\author{Cristiana Santos}
\email{c.teixeirasantos@uu.nl }
\affiliation{%
  \institution{Utrecht University}
  \country{The Netherlands}
}

\author{Nataliia Bielova}
\email{nataliia.bielova@inria.fr}
\author{Michael Toth}
\email{michael.toth@inria.fr}
\affiliation{%
  \institution{Inria}
  \country{France}
}

\author{Damian Clifford}
\email{damian.clifford@anu.edu.au}
\affiliation{%
  \institution{Australian National University}
  \country{Australia}
}

\renewcommand{\shortauthors}{Gray, et al.}

\begin{abstract}
 
User engagement with data privacy and security through consent banners has become a ubiquitous part of interacting with internet services. While previous work has addressed consent banners from either interaction design, legal, and ethics-focused perspectives, little research addresses the connections among multiple disciplinary approaches, including tensions and opportunities that transcend disciplinary boundaries. In this paper, we draw together perspectives and commentary from HCI, design, privacy and data protection, and legal research communities, using the language and strategies of ``dark patterns'' to perform an \textit{interaction criticism} reading of three different types of consent banners. Our analysis builds upon designer, interface, user, and social context lenses to raise tensions and synergies that arise together in complex, contingent, and conflicting ways in the act of designing consent banners. We conclude with opportunities for transdisciplinary dialogue across legal, ethical, computer science, and interactive systems scholarship to translate matters of ethical concern into public policy.

\end{abstract}

\ccsdesc[500]{Human-centered computing~User interface design}
\ccsdesc[500]{Social and professional topics~Governmental regulations}
\ccsdesc[500]{Social and professional topics~Codes of ethics}
\ccsdesc[500]{Security and privacy~Social aspects of security and privacy}

\keywords{Dark patterns, consent, GDPR, technology ethics, interaction criticism, transdisciplinarity}

\maketitle

\section{Introduction}
\label{sec:intro}
The language of ethics and values increasingly dominates both the academic discourse and the lived experiences of everyday users as they engage with designed technological systems and services. Within the HCI community, there has been a long history of engagement with ethical impact, including important revisions of the ACM Code of Ethics in the 1990s~\cite{Anderson1993-vu,Gotterbarn1998-la} and 2010s~\cite{Wolf2016-qg,Gotterbarn2017-nw,McNamara2018-yz}, development and propagation of ethics- and value-focused methods to encourage awareness of potential social impact~\cite{Detweiler2012-rw,Friedman2019-zg,Shilton2018-ws}, and the development of methodologies that seek to center the voices of citizens and everyday users~\cite{Bannon2018-kk,Olivier2015-dt,DiSalvo2014-ft}. In the past few years, everyday users have begun to become more aware of the ethical character of everyday technologies as well, with recent public calls to ban facial recognition technologies~\cite{ACM_US_Technology_Policy_Committee2020-ph} and further regulate privacy and data collection provisions~\cite{West2019-zd,Zaeem2020-ms}, alongside critiques and boycotts of major social media and technology companies by employees and users alike~\cite{Frenkel2020-bb,Suciu2020-ce}. These kinds of technology ethics issues have also been foregrounded by new and proposed laws and regulations---in particular,
the General Data Protection Regulation (GDPR) in the European Union~\cite{GDPR-16} and the California Consumer Privacy Act (CCPA) in the United States~\cite{CCPA-18}. These new legal standards have brought with them new opportunities to define unethical or unlawful design decisions, alongside new requirements that impact both industry stakeholders (e.g., `` data controllers'', ``data processors'', 
designers, developers) and end users.

Of course, HCI represents one of many disciplinary framings of technology ethics, with important parallel work occurring in other communities such as Science and Technology Studies (STS), Privacy, Ethics, and Law. As the ethical concerns present in technological systems and services become more apparent and widespread, others have called for a transdisciplinary engagement in conjunction with these other disciplinary perspectives to more fully address the complex intersections of technological affordances, user interactions, and near and far social impacts~\cite{Gray2019-ep,Kirkham2020-yn,Shilton2013-dq,Shilton2017-zu,Steen2015-qw}. Recent work has sought to bridge some of these perspectives through the use of \textit{dark patterns} as a theoretical framing, calling attention to a convergence of designer intent and negative user experience~\cite{Brignull2013-iu,Gray2018-or,Gray2020-zq,Greenberg2014-dg,Mirnig2017-ld,Mathur2019-hx,Nara-etal-20-ACMQueue}. We seek to explicitly build upon these traditions and concepts in this work.

In this paper, we draw together perspectives and commentary from HCI, design, privacy and data protection, and legal research communities, building an enhanced understanding of how these perspectives might arise together in complex, situated, contingent, and conflicting ways in the act of designing consent banners. 
The GDPR~\cite{GDPR-16} and the ePrivacy Directive~\cite{ePD-09} demand user consent for tracking technologies
and this requirement has resulted in a range of different techniques, interaction approaches, and even inclusion of dark patterns to gain user consent~\cite{Nouwens2020-ij,Soe2020-se,Matt-etal-19-SP,Grassl2020-lh}.
Thus, we took as our starting point for this paper a collection of consent processes for websites accessible to EU residents, where each consent process was captured through screen recording. We then built on prior analysis of this dataset to identify several consent patterns which were distributed across the temporal user experience that include initial framing, configuration, and acceptance of consent parameters. We then used our shared expertise as authors in HCI, design, ethics, computer science, and law to analyze these design outcomes for their legality using prior legal precedent~\cite{GDPR-16,ePD-09}, and their ethical appropriateness using relevant strategies from the dark patterns literature~\cite{Gray2018-or}. 

Our goal in this work is not to identify the breadth or depth of consent approaches, or even primarily to identify which of these approaches is most or least legally or ethically problematic. Instead, we use an \textit{interaction criticism}~\cite{Bardzell2011-hv} approach to analyze and reflect upon several common approaches to designing a consent banner from multiple perspectives: 1) design choices evident in the consenting process artifact itself; 2) the possible experience of the end user; 3) the possible intentions of the designer; and 4) the social milieu and impact of this milieu on the other three perspectives. Using this humanist approach to engaging with technological complexity, we are able to foreground conflicts based on role and perspective; identify how legal, design, and ethical guidance frequently conflicts or lacks enough guidance; and provide an interactive framework through which future work might assess ethical and legal impact across temporal aspects of the consenting process. 
This approach results in a detailed analysis of four consent strategies from multiple disciplinary and role-based perspectives, leading to an overview of the consent task flow in alignment with legal consent requirements and the identification of instances where dark patterns strategies are used to manipulate the user into selecting options that are not in their best interest.

The contribution of this paper is three-fold. First, we use a combination of legal and ethics frameworks to evaluate different approaches to obstructing or manipulating user choice when consenting, providing a range of examples to inform future policy work and ethics education. Second, we explore our exemplars using an interaction criticism approach, adding an ethics-focused layer to critical accounts of interactive systems. Third, we argue for transdisciplinary 
dialogue across legal, ethical, computer science, and interactive systems scholarship to translate matters of ethical concern into public policy.



\section{Related Work}\label{sec:related}

\subsection{Recent work on consent banners}
\label{sec:related-consent}

The most closely relevant work on which we build our contribution in this paper is a surge of studies on consent banners, including work primarily stemming from a legal compliance perspective \cite{Leenes2015TamingTC,Leenes2015TheC,Kosta2013PeekingIT,Santos2019-fe,Luguri2019ShiningAL} or a dark patterns or ``nudging'' perspective \cite{Nouwens2020-ij,Soe2020-se,Machuletz2020-nr,Utz-etal-19-CCS,Matt-etal-19-SP,Grassl2020-lh}. We will briefly summarize several key studies and findings in this area.


\subsubsection{Design choices that impact user behavior}
In 2019, Utz et al.~\cite{Utz-etal-19-CCS} conducted a field study on more than 80,000 German participants. Using a shopping website, they measured how the design of consent banners influence the behaviour of people acceptation or denial of consent. 
They found that small UI design decisions (such as changing the position of the notice from top to bottom of the screen) substantially impacts whether and how people interact with cookie consent notices. One of their experiments indicated that dark patterns strategies such as \emph{interface interference} (highlighting ``Accept'' button in a binary choice with ``Decline''), and \emph{pre-selected choices} for different uses of cookies has a strong impact of whether the users accept the third-party cookies.


In their 2020 study, Nouwens et al.~\cite{Nouwens2020-ij} performed a study on the impact of various design choices relating to consent notices, user interface nudges and the level of granularity of options. 
They scraped the design and text of the five most popular CMPs on top 10,000 websites in the UK, 
looking for the presence of three features: 1) if the consent was given in an explicit or implicit form; 2) whether the ease of acceptance was the same as rejection---by checking whether accept is the same widget (on the same hierarchy) as reject; and 3) if the banner contained pre-ticked boxes, considered as non-compliant under the GDPR~\cite[Recital 32]{GDPR-16}. In their results, they found less than 12\% of the websites they analyzed to be compliant with EU law. In their second experiment, they ran a user study on 40 participants, looking at the effect of 8 specific design on users' consent choices. They recorded an increase of 22 percentage points in given consent when the ``Reject all'' button is removed from the first page, and ``hidden'' at least two clicks away from this first page. Finally, they found a decrease of 8 to 20 percentage points when the control options are placed on the first page.

Machuletz and Böhme~\cite{Machuletz2020-nr} set up a user study of post-GDPR consent banners with 150 students in Germany and Austria. Building upon with behavioural theories in psychology and consumer research, they evaluated the impacts of 1) the number of options displayed to the user, and 2) the presence/absence of a ``Select all'' default button in the banners, nudging the user toward giving a complete consent. They showed a significant increase in consent when the highlighted default ``Select all'' button is present, with participants often expressing regret about their choice after the experiment.

Soe et al.~\cite{Soe2020-se} performed a manual analysis of GDPR-related consent banners. They manually collected banners from 300 Scandinavian and English-speaking news services, looking for manipulative strategies potentially circumventing the requirements of the GDPR. Then, they analyzed the design of these banners, and ``found that all employ some level of unethical practices''. In their findings, the most common patterns were \emph{obstruction}, present in 43\% of the tested websites containing dark patterns, and \emph{interface interference}, present in 45.3\%.

\subsubsection{Issues with compliance and detection}

In their 2019 study, Matte et al.~\cite{Matt-etal-19-SP} focused on Consent Management Platforms (CMPs) implementing IAB Europe's Transparency and Consent Framework (TCF) framework. They analyzed consent stored behind the user interface of TCF consent banners. They detected suspected violations of the GDPR and ePrivacy Directive by running two automatic and semi-automatic crawl campaigns, on a total of 28,257 EU websites. Specifically, they studied 1) whether consent was stored before the user made the choice, 2) whether the notice offers a way to opt out, 3) whether there were pre-selected choices, and 4) if the choice that the user had made was respected at all. They found 141 websites registering positive consent before the user's choice, 236 websites that nudged users towards accepting consent by pre-selecting options, and 27 websites that storing a (false) positive consent even if the user had explicitly opted out. They also developed free and open-source tools to enable DPAs and regular users to verify if consent stored by CMPs corresponds to their choice.

Human and Cech~\cite{Human2021-xz} built a theoretical framework to evaluate consent collection from five major tech companies---Google, Amazon, Facebook, Apple, and Microsoft---focusing on interactions, graphical design, and text. They noticed asymmetric design, hidden information, and unclear statements. They show the way these companies gather consent to be ethically problematic, and sometimes non GDPR-compliant.
%

Finally, Santos et al.~\cite{Santos2019-fe} performed an interdisciplinary analysis of the legal and technical requirements of consent banners under the GDPR and ePD, identifying 22 requirements from legal sources and both technical and legal experts to verify compliance of consent banner design. They explored ways to realize manual or automated verification of these requirements, aiming to help regulators, NGOs, and other researchers to detect violation of EU legislation in consent banner implementation. They also showed which requirements are impossible to verify in the current web architecture.

\emph{Summary.}
Prior work has evaluated the impact of interface design on consent banners and the decisions of users. These studies have primarily addressed: a) the computational detection of concrete design choices evident in source code; b) the user impact of these design choices; and c) the legitimacy of some of these design choices from a ethics, legal, or policy perspective. However, much of this work has occurred in silos, resulting in a disconnection of these design choices from the overall consent flow, or a lack of identification of the ways in which particular dark patterns might be connected to legal requirements and the user experience. In this paper, we build upon this gap in substantive transdisciplinary discourse, addressing a cross-section of legal, design, and technical expertise in relation to consent design choices and dark patterns. 



\subsection{Practitioner- and Academic-Focused Discussions of Ethics}
\label{sec:related-practitionner}

Previous scholarship has revealed markedly different discourses regarding ethical concerns, with the academic community largely focused on arguing in relation to moral and ethics theory (e.g., \cite{Friedman2019-zg,Friedman2002-jm,Shilton2018-ws}) and the practitioner community focused more on tangible and problematic practices (e.g., \cite{Bosch2016-vc,Gray2019-ep,Gray2020-zq,Brignull2013-iu}). While there has been substantial interest in ethically-focused design practices in the HCI community for decades, most of this work has been subsumed into one of three categories: 1) the development and maintenance of a code of ethics in the ACM, including relevant use of this code in education and practice \cite{Wolf2016-qg,Gotterbarn2017-nw,McNamara2018-yz};  2) the construction and validation of methods to support ethics-focused practice, most commonly within the methodology of Value-Sensitive Design (VSD; \cite{Friedman2002-jm,Friedman2019-zg}); and 3) the use of practitioner-focused research to reveal patterns of ethical awareness and complexity \cite{Gray2019-ep,Chivukula2020-bv,Watkins2020-zr,Shilton2018-sw,Shilton2017-zu,Shilton2013-dq,Shilton2014-fx,Steen2015-qw}. Work on VSD has also included efforts across these categories that identify opportunities for implementation in design and evaluation activities~\cite{Cummings2006-nt,Van_Wynsberghe2013-hf,Shilton2017-zu} as well as broader engagement in ethics-focused argumentation, building connections from ethical and moral theories to HCI and Science and Technology Studies (STS) concerns (e.g., \cite{Borning2012-re,Manders-Huits2011-dt,Davis2021-yt,Jacobs2018-hz,Le_Dantec2009-av}). One particular source of interest that relates to the framing of this paper is a recent paper by Kirkham \cite{Kirkham2020-yn} that links ethical concerns with VSD and guidance from the European Convention on Human Rights; \cite{Kirkham2020-yn} is one of few examples of legal, ethical, and HCI discourses coming together with the goal of informing HCI scholarship and guidance that may inform design practices.

The practitioner discourse regarding ethics has been more diffuse, representing an interest in ethics-focused work practices (e.g., Nodder's \textit{Evil by Design} \cite{Nodder2013-rw}), but perhaps the most vital conversations have emerged around the conceptual language of ``dark patterns.'' This term was coined by Harry Brignull in 2010 to describe ``a user interface carefully crafted to trick users into doing things they might not otherwise do [\ldots] with a solid understanding of human psychology, and [which] do not have the user's interests in mind''~\cite{Brignull2013-iu}. Brignull identified a taxonomy~\cite{Brignull2015-il} of twelve different types of dark patterns and collects examples in his ``hall of shame,'' which has subsequently been built upon by Gray et al.~\cite{Gray2018-or}, B\"{o}sch et al.~\cite{Bosch2016-vc}, and Mathur et al.~\cite{Mathur2019-hx}. In 2016, B\"{o}sch et al. presented a classification of eight ``dark strategies''~\cite{Bosch2016-vc}, built in opposition to Hoepman's ``privacy design strategies''~\cite{Hoep-2013-PLSC}, which uncovered several new patterns: \emph{Privacy Zuckering}, \emph{Bad Defaults}; \emph{Forced Registration} (requiring account registration to access some functionality); \emph{Hidden Legalese Stipulations} (hiding malicious information in lengthy terms and conditions); \emph{Immortal Accounts}; \emph{Address Book Leeching}; and \emph{Shadow User Profiles}. These patterns were later extended in an online privacy dark pattern portal~\cite{Bosch-website} for the community to study and discuss existing patterns and contribute new ones. Mathur et al.~\cite{Mathur2019-hx} used automated techniques to detect text-based dark patterns, such as \emph{framing}, in a set of \textasciitilde53K product pages from \textasciitilde11K shopping websites. They found 1,818 occurrences of dark patterns, involving 183 websites and 22 third-party entities. They built a classification of these dark patterns, dividing them in 15 types and 7 categories, and a taxonomy of their characteristics. Finally, they made some recommendations to mitigate the negative effects of these deceptive techniques on users. In this work, we rely more specifically on the five dark patterns strategies proposed by Gray et al.~\cite{Gray2018-or}, which include: \textit{nagging}--a ``redirection of expected functionality that persists beyond one or more interactions''; \textit{obstruction}--''making a process more difficult than it needs to be, with the intent of dissuading certain action(s)''; \textit{sneaking}--``attempting to hide, disguise, or delay the divulging of information that is relevant to the user''; \textit{interface interference}--``manipulation of the user interface that privileges certain actions over others''; and \textit{forced action}--``requiring the user to perform a certain action to access (or continue to access) certain functionality.''

In other complementary work addressing dark patterns, scholars have described how dark patterns are perceived from an end-user perspective \cite{Maier2020-yk,Gray2020-ei}, how these patterns appear in non-screen-based proxemic interactions \cite{Greenberg2014-dg} and in mobile interactions \cite{Di_Geronimo2020-mh}, how these patterns can impact online disclosure \cite{Waldman2019-km}, and how these patterns can be used to motivate design discourses and argumentation about ethics \cite{Dieter2015-wl}. Finally, Chivukula and Gray \cite{Chivukula2019-yb,Gray2020-zq} have recently shown how interest in dark patterns can reveal larger patterns of coercion and abuse in digital technologies, building on the popular subreddit ``r/assholedesign'' to define properties of an ``asshole designer.''

 {\em Summary.} Previous work relating to ethics addresses a broad range of concerns, arguing from moral philosophy and professional ethics, engaging with complexity from the practitioner perspective, or some combination of these perspectives. We seek to connect these concerns in a transdisciplinary framing, better connecting practitioner and academic concerns about ethics within the context of legal and design concerns using the language of ``dark patterns.''

\subsection{Legal scholarship on cookie banners and consent requirements}
While legal scholarship infrequently intersects with work from the HCI community (see \cite{Kirkham2020-yn,Spaa2019-vm} for rare examples connecting HCI to policymaking), literature from a legal perspective is vital to our understanding of what practices may be lawful or unlawful, and how these policies emerge and are then tested by the courts. To provide a basis for arguing from a legal perspective in this paper, we provide a brief summary of some of the key legislation and requirements dictated by GDPR, which ground our analysis of problematic consent banners in Section~\ref{sec:findings}.

\label{sec:related-legal}
GDPR is the key pillar of the EU data protection framework, as supplemented by the ePrivacy Directive. In essence, the regulation formulates standards for the processing of personal data. 
Personal data are defined  as \emph{``[\ldots] any information relating to an identified or identifiable natural person (‘data subject’) [\ldots]''} (Article 4(1) GDPR). 
Processing is similarly broadly defined and amounts to any action undertaken with such information (Article 4(2)).
The GDPR regulates the processing of personal data by affording rights to users (called ``data subjects''), by imposing obligations of entities that process personal data (data controllers and processors), and a monitoring role for data protection authorities (DPAs). 

GDPR introduces specific {\em principles relating to the processing of personal data} (Article 5 GDPR) which guide data controllers and processors in the interpretation of the rights and obligations.
Of immediate importance for are the lawfulness, fairness and transparency principles (Articles 5(1)(a) GDPR) and the accountability principle (Article 5(2) GDPR). 
The processing of personal data requires one of the conditions for lawful processing to be satisfied (Article 6(1) GDPR), namely {\em consent}.
%
%
The ePrivacy Directive  stipulates that user consent is required for processing information through the use of tracking technologies (which includes cookies, (Article 5(3) of the ePrivacy Directive)\footnote{Only functional cookies which are used for communications and strictly necessary purposes are exempted of consent. See more detailed analysis of the scope of consent in Santos et al.~\cite[Section 4]{Santos2019-fe}.}.
Consent is commonly expressed through interface design elements in the form of a pop-up.

 Table~\ref{tab:designs-patterns-consent} presents a synthesis of the legal requirements for valid consent which stem from the GDPR, the ePrivacy Directive (ePD) and the Court of Justice of the EU (CJEU).
%
Consent is defined in Article 4(11) and complemented by Articles 6 and 7 of the GDPR which states 
that for consent to be valid, it must satisfy the following elements: it must be ``freely given, specific, informed and unambiguous.'' 
The controller is required to be able to demonstrate consent (Article 7(1) GDPR) keeping in mind that, in assessing the ``freely given'' definitional condition, rendering access to the service conditional on consent may invalidate the reliance on consent (Article 7(4) GDPR). In short, consent is required to be presented in a manner which is clearly distinguishable from other matters (Article 7(2) GDPR) and represent a meaningful choice as evidenced by the ability to withdraw consent (Article 7(3) GDPR). 

{\em Summary.}
Legal scholarship has not yet provided a threshold for the appropriateness of specific design patterns in consent banners, and which requirements for a valid consent are or are not respected in each case. Although some regulators have provided classifications of dark patterns applied to various practices which have been deemed unfriendly in terms of privacy impacts, these classifications have not qualified which dark patterns are potentially unlawful and which legal requirements are potentially violated in relation to these patterns. We seek to address this gap, focusing on a legal compliance perspective by analyzing the lawfulness of these dark patterns from the consent requirements side.

\begin{table*}[]
\centering
\begin{tabular}{p{3.5 cm} p {6 cm} p {7.4 cm}} 
\toprule
  \textbf{Requirements} & \textbf{Provenance in the GDPR, ePD, CJEU}  & \textbf{Description} \\ \midrule
 
 
Freely given & Art. 4(11), 7(4) GDPR & Consent should imply a voluntary choice to accept/decline the processing of personal data, taken in the absence of any kind of pressure or compulsion on the user \vspace{5px} \\ 
 
Specific & Art. 4(11) GDPR, CJEU Planet 49~\cite{CJEU-Planet49-19} & Consent should be separately requested for each purpose \vspace{5px} \\ 
 
Informed & Art. 4(11) GDPR, 5(3) ePD, CJEU Planet 49~\cite{CJEU-Planet49-19} & The user must be given clear and comprehensive information about what data is processed, the purposes and means for expressing consent \vspace{5px} \\ 

Unambiguous & Art. 4(11) GDPR, CJEU Planet 49~\cite{CJEU-Planet49-19} & Clear and affirmative action of the user \vspace{5px} \\ 

Readable and accessible & Art. 7(2), Recitals 32, 42 GDPR & Consent request should be distinguishable of other matters, intelligible, accessible to the user, using clear and plain language, not unnecessarily disruptive to the use of the website \\ 

\bottomrule
\end{tabular}
\caption{Legal requirements for a valid consent, provenance in the GDPR, ePrivacy Directive (ePD) and the Court of Justice of the EU (CJEU).}
\label{tab:designs-patterns-consent}
\end{table*}

\section{Our Approach}
\label{sec:approach}



\subsection{Researcher Positionality}
\label{sec:approach-positionality}

We explicitly and intentionally framed this project---and our broader research collaboration---in relation to transdisciplinary scholarship that expands beyond any one of the authors' respective disciplines. As one effort to acknowledge the subjective positions from which our readings of each consent banner emerges, we include a brief description of our disciplinary expertise as a means of increasing the transparency of our research efforts~\cite{Patton2001-mz}.


The authors of this paper are researchers that engage in research, design, or development across the following domains:
\begin{itemize}
    \item Bielova and Toth are computer scientists with expertise in web privacy measurement and privacy compliance;
    \item Gray is an HCI and design researcher with expertise in UX, ethics, values, and dark patterns;
    \item Santos and Clifford are legal scholars with particular expertise in EU Data Protection law.
\end{itemize}

These different areas of disciplinary expertise are frequently contested, working in silos, or are otherwise in conflict with concepts or guidance from other disciplinary perspectives. We use the concept of ``dark patterns'' as a primary example of our means of connection to each other as scholars, while also recognizing that the concept of dark patterns has been addressed separately within the research communities of HCI, Computer Science and Law, with varying degrees of impact and limited interdisciplinary effort. In this paper, we explicitly leverage our collective attempts as a research team to bridge disciplinary silos as a way of collectively discussing future transdisciplinary approaches to ethics, policy, design, and computer science. This paper was written over a period of almost nine months, involving numerous online calls where we engaged with the transdisciplinary complexity of this space, seeking both to find a ``common ground''---where concepts from each of our disciplinary perspectives might find resonance---as well as identifying how the emergent findings that are present in our argumentation might point towards disciplinary advances in each of our respective areas, and how these might be productively brought together as an example of transdisciplinary scholarship for the HCI community. 

\subsection{Data Collection and Framing}
\label{sec:approach-collection}

Due to the argumentation focus of this paper, we relied upon data sources collected in previous projects to identify salient consent design choices to elaborate further. From 2019--2020, a subset of the authors collected a broad range of examples of consent banners, using screen recording software to capture the entire interaction flow required to fully consent in accordance with GDPR requirements.
The screen recordings were made using desktop-class devices only, recognizing that mobile experiences themselves are an important space for future work, likely with different forms of pattern instantiation and sources of manipulation. 

%

\begin{figure}[ht]
 \centering
  \includegraphics[width=\linewidth]{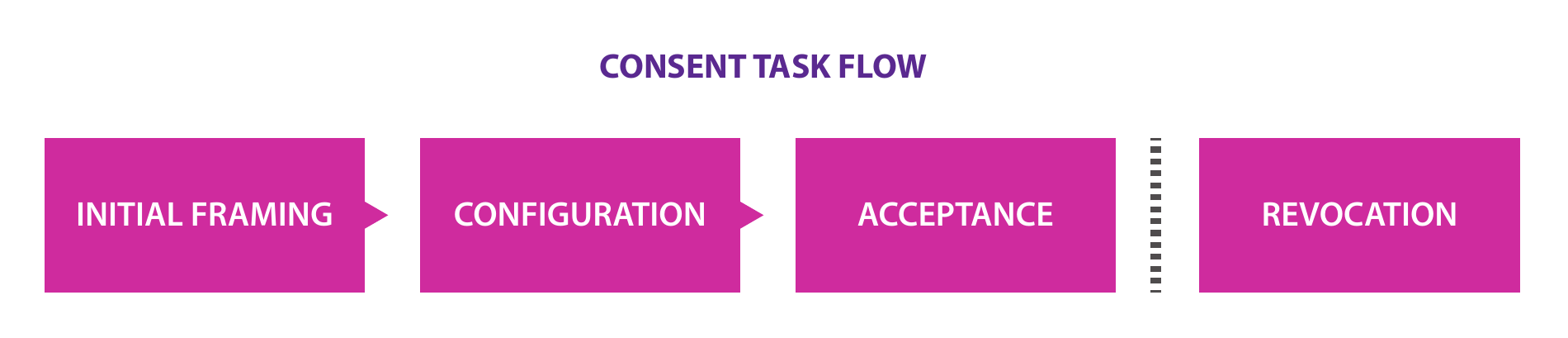}
 \caption{The task flow of the consenting process by phase.}
 \Description[Task flow: 1) Initial framing > 2) Configuration > 3) Acceptance | 4) Revocation]{4 boxes summarizing the user's task flow in a consent banner: 1) Initial framing > 2) Configuration > 3) Acceptance | 4) Revocation}
 \label{fig:config}
\end{figure}

We based our analysis on a dataset of 560 websites accessible from the EU from French-, Italian- or English- speaking countries: France, UK, Belgium, Ireland and Italy, and \texttt{.com} websites from Matte et al.~\cite{Matt-etal-19-SP}. Each of these sites was detected automatically in this prior work as containing a consent banner that implemented IAB Europe Transparency and Consent Framework (TCF~\cite{IAB_Europe2019-bz}). These 560 websites also belonged to 1,000 top Tranco~\cite{Le_Pochat2018-xt} list, which indicates popular websites of the top level domain (TLD) of the above-mentioned European countries (e.g., \texttt{.fr}, \texttt{.uk}, etc) and domain \texttt{.com}. 
%
From this dataset, 
we focused on locating a range of potentially manipulative design exemplars, using recorded videos or screenshots of the consent experiences to support a manual and collaborative analysis of their design and text.
In total, we reviewed recordings from over 50 sites 
and extensively analyzed the design and users' means of interaction with the consent banners on these websites.
While reviewing other recent and relevant literature on ethical issues in the design of consent banners (e.g., \cite{Nouwens2020-ij,Machuletz2020-nr,Soe2020-se}), we identified four main phases in the consent task flow (Figure~\ref{fig:config}): 
%
%
\begin{enumerate}
\item the initial framing as a user enters the site; 
\item the presentation of configuration options to accept or select more precise consent options; 
\item the means of accepting the configuration options; and 
\item the ability to ultimately revoke consent. 
\end{enumerate}

Within this task flow, we worked as a research team to identify four different combinations of design choices that were represented in the dataset and raised productive ethical dilemmas when viewed from multiple disciplinary perspectives. For instance, the consent types \emph{reduced service} and \emph{consent wall} we describe in the findings section of this paper had not previously been detected or precisely identified in prior work, either through empirical web measurements or from user studies. Moreover, EU Data Protection Authorities currently have conflicting opinions on the lawfulness of these practices that this analysis can clarify to policy makers~\cite{Santos2019-fe}; these opinions can only be derived from a legal, design, and computer science analysis of consent requirements and consent banners as we set out in the remainder of this paper.

Because our main goal in this paper is to examine the complexity of these design outcomes and not to identify how common these patterns occur ``in the wild,'' we used the dataset as a source of inspiration and departure rather than as a means of conducting a content analysis or other inductive form of inquiry.
We therefore focused on analyzing specific \textit{types} of banners with the goal of representing a broad range of consenting approaches through our interdisciplinary perspectives rather than what was most typical or used on the most popular sites.


\subsection{Data Analysis}
\label{sec:approach-analysis}

Within each element of the task flow as embodied by a specific set of design choices, we inspected specific forms of manipulation through our analysis. We used the practice of \textit{interaction criticism} \cite{Bardzell2011-hv} to investigate and interrogate manipulation from multiple perspectives. According to Bardzell~\cite{Bardzell2011-hv}, the practice of \textit{interaction criticism} is the ``rigorous interpretive interrogations of the complex relationships between (a) the interface, including its material and perceptual qualities as well as its broader situatedness in visual languages and culture and (b) the user experience, including the meanings, behaviors, perceptions, affects, insights, and social sensibilities that arise in the context of interaction and its outcomes.'' The process of engaging in criticism builds upon four perspectives or positions of argumentation: 1) the designer; 2) the interface itself; 3) the user; and 4) the social context of creation and use. In our work, we intend to build upon the practice of interaction criticism by highlighting the contributions of design scholarship, legal jurisprudence, and discussions of ethics and values from both academic and practitioner perspectives. Using this approach provided a conceptual means by which we could each intentionally de-center our own disciplinary expertise and vocabulary, foregrounding perspectives and concepts from other disciplinary traditions and subject positions in the search for common ground. Across these disciplinary perspectives, we sought to include a number of potential considerations:
 \begin{enumerate}
    \item the designer's potential intent in relation to the design choice; \\ \textit{potential considerations include: design judgments, context- or role-based limitations of the designer's work, means of balancing multiple constraints, use of design precedent}
    \item the designed interface itself; \\ \textit{potential considerations include: formal aspects of the UI, common design patterns that are exemplified by the interface under evaluation, indications of designed interactions or user experience inscribed into the interface, language used, typographic and compositional decisions, indication of feedforward}
    \item the perspective and experience of the end user; \\ \textit{potential considerations include: anticipated user interactions and experience, technical knowledge required or assumed of the end user, designer's perception of the system model} 
    \item the potential social impact of the designed experience. \\ \textit{potential considerations include: relevant business models and economic rationale, current and future role of technology, social acceptance or rejection of technology norms, agency of users and technology providers}
 \end{enumerate} 

Using this approach, we iteratively built out an argument from each of the perspectives listed above, seeking to identify salient design principles, potential social expectations or means of describing intent, and legal or policy guidance through which the consent design choices could be framed. Through this process, authors with expertise across a range of disciplinary perspectives added their own sources of evidence, while also reviewing the coherence of argumentation from other disciplinary perspectives. We used the qualitative/interpretivist notion of reflexivity to continuously identify strengths and gaps, seeking not to reach objective and final consensus, but rather to explore differences in disciplinary perspectives and the points at which these perspectives overlapped or collided.

\section{Findings}
\label{sec:findings}

We organize our findings based on the temporal direction of a user's task flow, investigating four design choices in relation to the consenting task flow. Revocation is the fourth element of the task flow, which we include in Figure~\ref{fig:config}; however, we do not address revocation in our analysis approach. Across these user consent tasks and criticism perspectives, we engage in an interaction criticism analysis over the following sections, particularly highlighting the interplay of legal requirements, potential violations, and possible gaps in legal and policy guidance. In Figure~\ref{fig:flowchart} we summarize how this set of design choices relates to legal requirements and dark patterns strategies in the context of the overall consent task flow.



\subsection{Initial Framing}
\label{sec:findings-initial}

The ``initial framing,'' according to Figure~\ref{fig:config}, corresponds to the very first component of the consent mechanism a user sees when entering a website. This framing typically consists of an information banner disclosing the tracking technologies used and their purposes for data processing, with an acceptance button, and a link to the website's privacy policy. The initial framing banner can also take the form of a dialog or popover displayed on a part of the page, but may also completely block the page, preventing any action by the user until a choice has is made, such as the consent wall and tracking wall types that we analyze in the following sections.

\subsubsection{Consent Wall}
\label{sec:findings-consent-wall}

\begin{figure*}[ht]
  \centering
  \includegraphics[width=\linewidth]{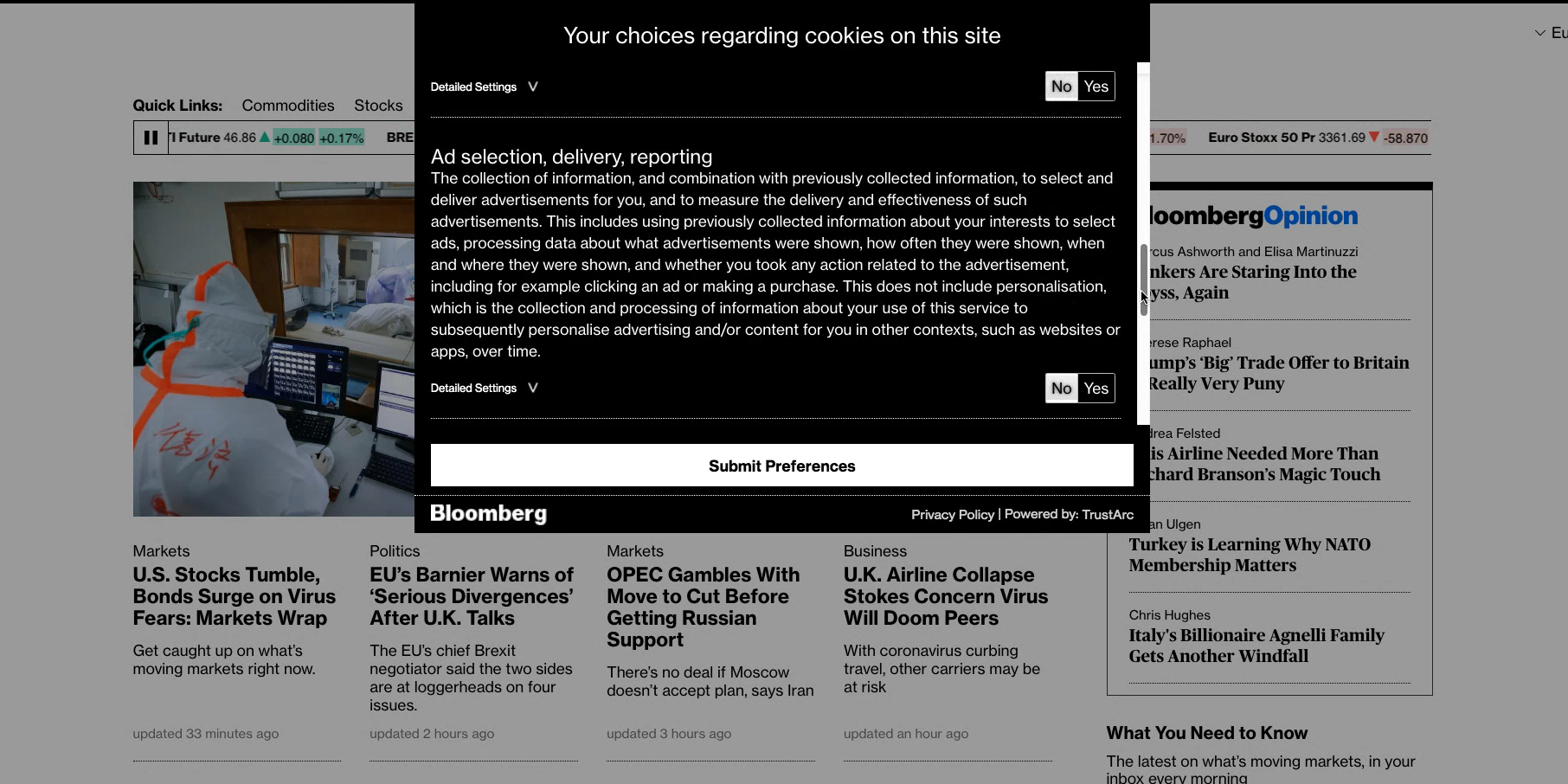}
  \caption{An example of a consent wall, recorded on 5 March 2020. Credits: Bloomberg.com}
  \Description[Screenshot of Bloomberg's website with the banner blocking access]{Screenshot of Bloomberg's website showing a consent banner blocking access to the user unless they make a choice}
  \label{fig:ConsentWall}
\end{figure*}

A consent wall is a design choice that blocks access to the website until a the user expresses their choice regarding consent. 
This design choice allows a user to select between acceptance and refusal; however, the concrete use of the website is blocked until a choice has been made. An example from the website of \url{https://www.bloomberg.com/europe} illustrates the use of a consent banner forcing the user to make a choice, thus blocking the access of the website, as shown in Figure~\ref{fig:ConsentWall}\footnote{Video recorded on 5 March 2020: \url{https://mybox.inria.fr/f/28f689abbd8a4f188c89/}}.

We now consider this design choice from four different perspectives, in line with the interaction criticism perspective, overlaying our analysis with legal analysis and commentary regarding the implementation of dark patterns. 

\paragraph{Designer perspective}
The use of a consent wall clearly separates the consent process from the use of the underlying website, visually and also interactively. Thus, the actual presentation of the consent banner relies upon typical, while perhaps not fully ethical, interface design patterns.  If from the \emph{designer perspective}, the intent is likely to be read as reducing user choice through the layering and locking out of functionality, it could be deemed to be both manipulative and coercive. However, from a \textit{legal perspective},
including a visual limitation---such as blocking access to a website until a user expresses a choice---will force the user to consent and therefore it possibly violates a \textit{freely given} consent. 
Rendering access to a service conditionally based upon consent could raise serious concerns in relation to the ‘freely given’ stipulation in the definition of consent (Article 7(4) of the GDPR as further specified in Recital 43 thereof). In effect, one could take the view that this separation between a ``consent request'' \textit{vis \`a vis} ``content of the request'' could be considered as being of strategic advantage to the designer, since the separation of these codebases---one mandatory and site-wide impacting the entire user experience, and another that is page-specific---might naturally lead to decisions such as a consent wall. Additionally, the designer/controller is required to be able to demonstrate consent (Article 7(1) GDPR) and to fulfill this requirement, consent must be presented in a manner which is clearly distinguishable from other matters, which may potentially support aspects of such interruption to the user experience and serve as a practical implementation of the obligations laid down by the Regulation. Although, we observe that such an interpretation would appear to subside in the face of a teleological interpretation of the GDPR. If the user's choice does not correspond to the expected choices of the website publisher/designer, the website should provide other means of accessing the same version of the website (such as paid options), where the user's choice is respected. A consent wall that blocks the service provided by the website without other options has detrimental effects.

\paragraph{User perspective}
A \textit{user} demonstrates their intent to gain access to the content of a site by navigating to a particular URL or by clicking a link. This intent points to their desire to access the content of a website, and only after the site loads do they face obstructive overlays that are of secondary importance to many users beyond the content the user was intentionally navigating towards. In this way, the consent wall can be considered as a visual and interactive barrier to desired content, exemplifying the dark pattern strategy of \textit{forced action,} defined as ``requiring the user to perform a certain action to access [\ldots] certain functionality.'' In addition, although this design choice allows some degree of accessibility or interaction, a consent wall
could also be considered as an \textit{obstruction} to the user's primary intention to access the full content of the website visited, 
with the relative weight or impact of this dark pattern of \emph{obstruction} to be based on the amount of content or interactivity that is obscured or limited. 

\paragraph{Interface perspective}
The manipulation evident in the designed interface is intentionally structured to achieve a higher collected number of positive consents from users through the use of layered strategic elements as popovers, lightboxed forms, and other means of layering content to encourage consent---and by comparison, discourage rejection. Notably, when content desired and deemed relevant to the user is not made immediately or readily accessible, and is instead hidden under an overlay or other interface elements, with the primary motivator \emph{to disguise relevant information as irrelevant}, the interface decisions could also point towards the use of \textit{obstruction} in placing visual and interactive barriers between the target of the user's interaction (the content) and the only salient interactive target provided by the site (the consent banner). 


This design choice may violate another requirement named the ``readable and accessible consent request'' (Article 7(2)), meaning that a consent request should not be unnecessarily disruptive to the use of the service for which it is provided (Recital 32).
Thus, it could be argued that consent walls are confusing and unnecessarily disruptive of the user experience, and other consent design implementations could be sought while engaging users. This legal evaluation of the interface decisions requires a more evidence-based assessment of what will amount to a concrete implementation of what is ``unnecessarily disruptive.'' In fact, much depends on the context at hand---as experienced by the end user---but with these provisions in mind, it could be argued that although a consent wall may be a legitimate means of requesting consent, the user should also have the flexibility to cancel the request and continue browsing without the burden of tracking. 
Practically speaking therefore, compliance with this legal requirement of a freely given consent is \emph{context dependent}.

\paragraph{Social impact perspective}
Positioning consent as the main mechanism to access desirable content could result in consent auto-acceptance or consent fatigue, where users tend to automatically dismiss 
any selection options in their path in order to achieve their goal. And it is this potential 
that demonstrates how the legitimacy of consent walls---from a legal perspective---is a complicated question. Across multiple websites, the immediate request for consent could take on the characteristics of the dark pattern ``nagging''---which while not inherently harmful as a single case, gains strength through its ubiquity across multiple web experiences that may be experienced in a single web browsing session. Thus, the social relationship shown to be valued through the GDPR is one where the interruption of service may be seen as useful, or even necessary depending on the context at stake.

\paragraph{Summary} This design choice presents a tension between i. interactive separation of user activities, ii. strategies meant to limit user interaction prior to completing the consenting process, iii. requirements that mandate that consenting precedes use, and iv. the various impacts of both a ``burden of care'' on the part of the designer and the ``freely given'' nature of the consent process itself. These tensions, while potentially pointing towards rejection of this design choice as legally acceptable, also show the diminished user experience and unnecessary fragmentation of the user experience in order to satisfy legal requirements.


\subsubsection{Tracking Wall}
\label{sec:findings-tracking-wall}

\begin{figure}[ht]
\centering
\begin{subfigure}[t]{0.45\textwidth}
  \includegraphics[width=\linewidth]{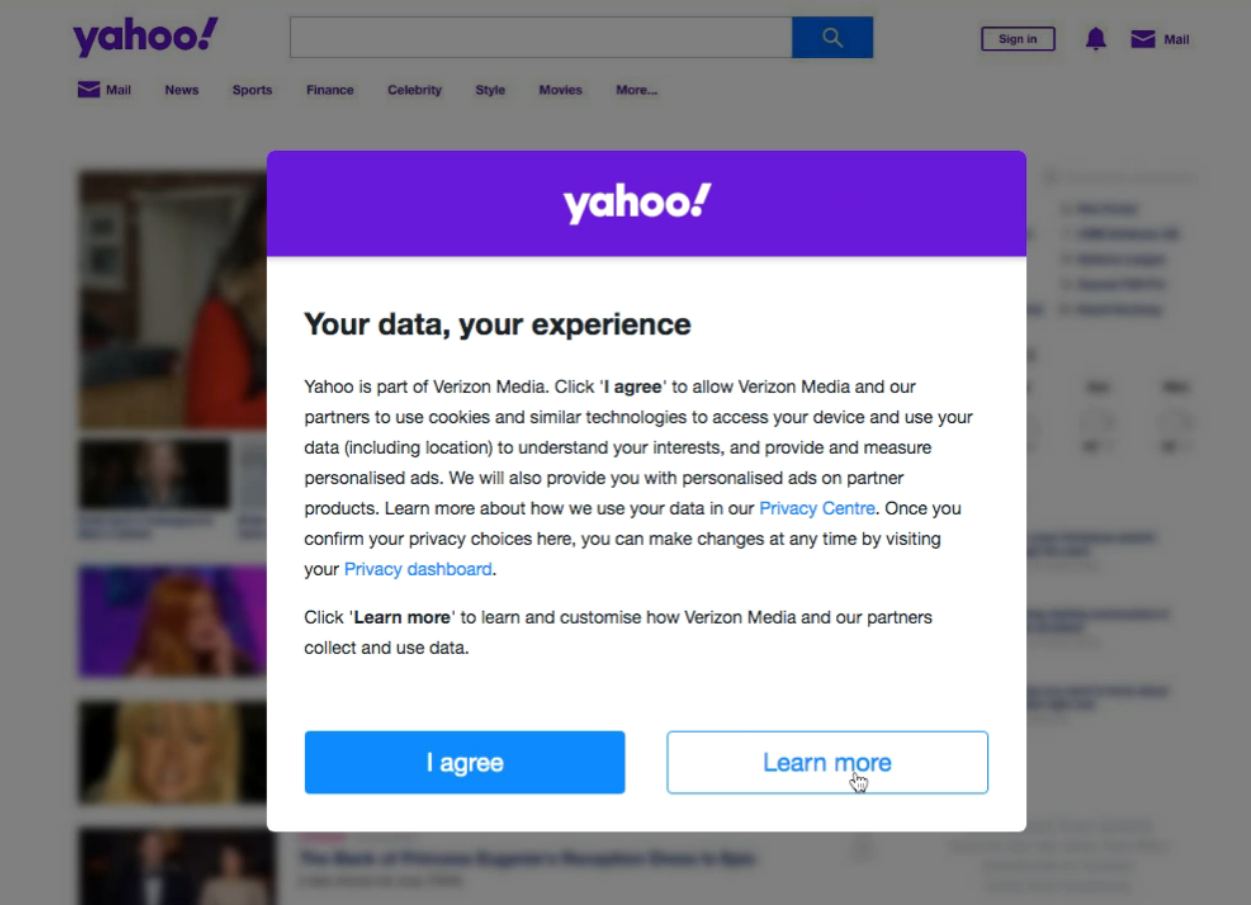}
  \caption{Initial framing}
  \Description[Screenshot of Yahoo's website with the banner blocking access]{Screenshot of Yahoo's website with the banner blocking access, with only the choices "I agree" or "Learn more" available to the user}
\end{subfigure}\hfill
\begin{subfigure}[t]{0.45\textwidth}
  \includegraphics[width=\linewidth]{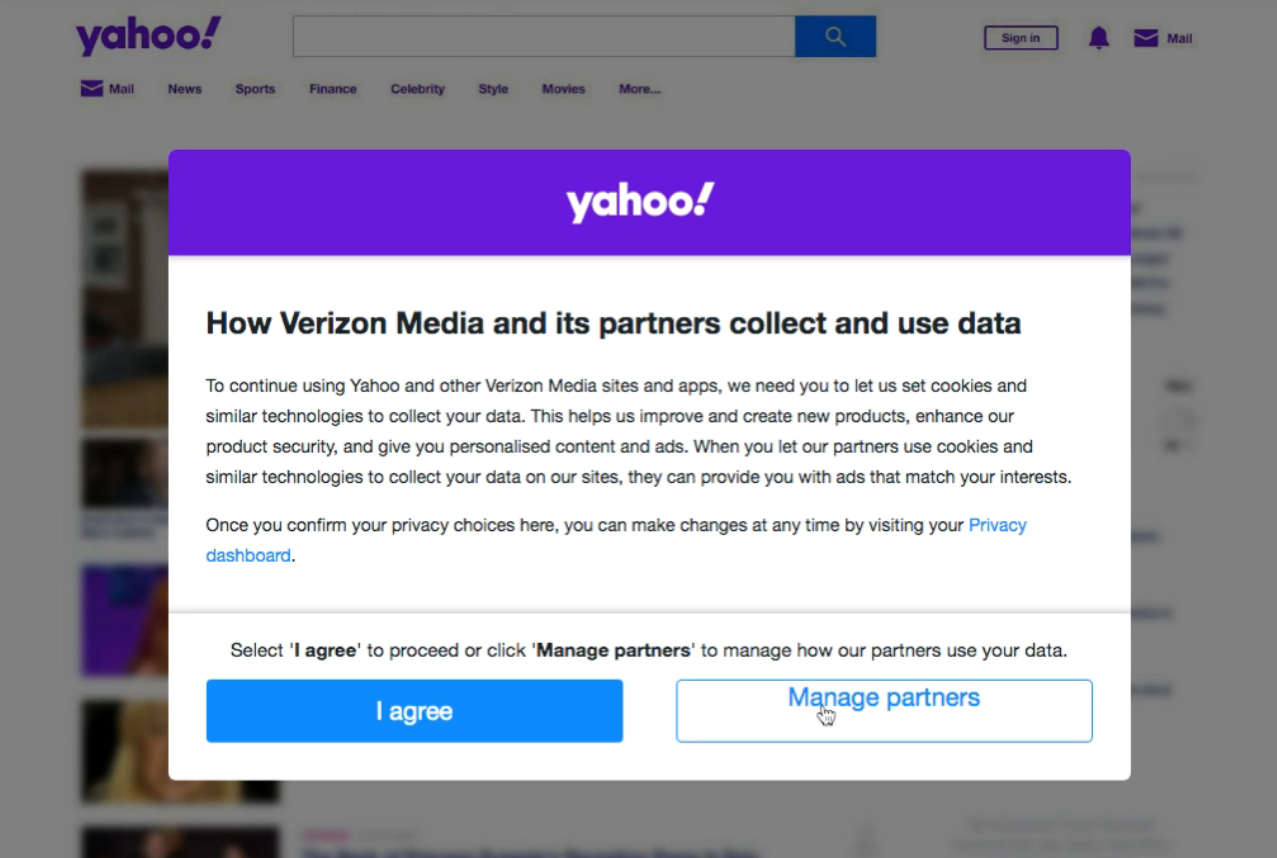}
  \caption{Short privacy policy}
  \Description[Screenshot of Yahoo's website with a short privacy policy present]{Screenshot of Yahoo's website with a short privacy policy present, along with buttons for "I agree" or "Manage partners"}
\end{subfigure}\hfill
\begin{subfigure}[t]{0.45\textwidth}
  \includegraphics[width=\linewidth]{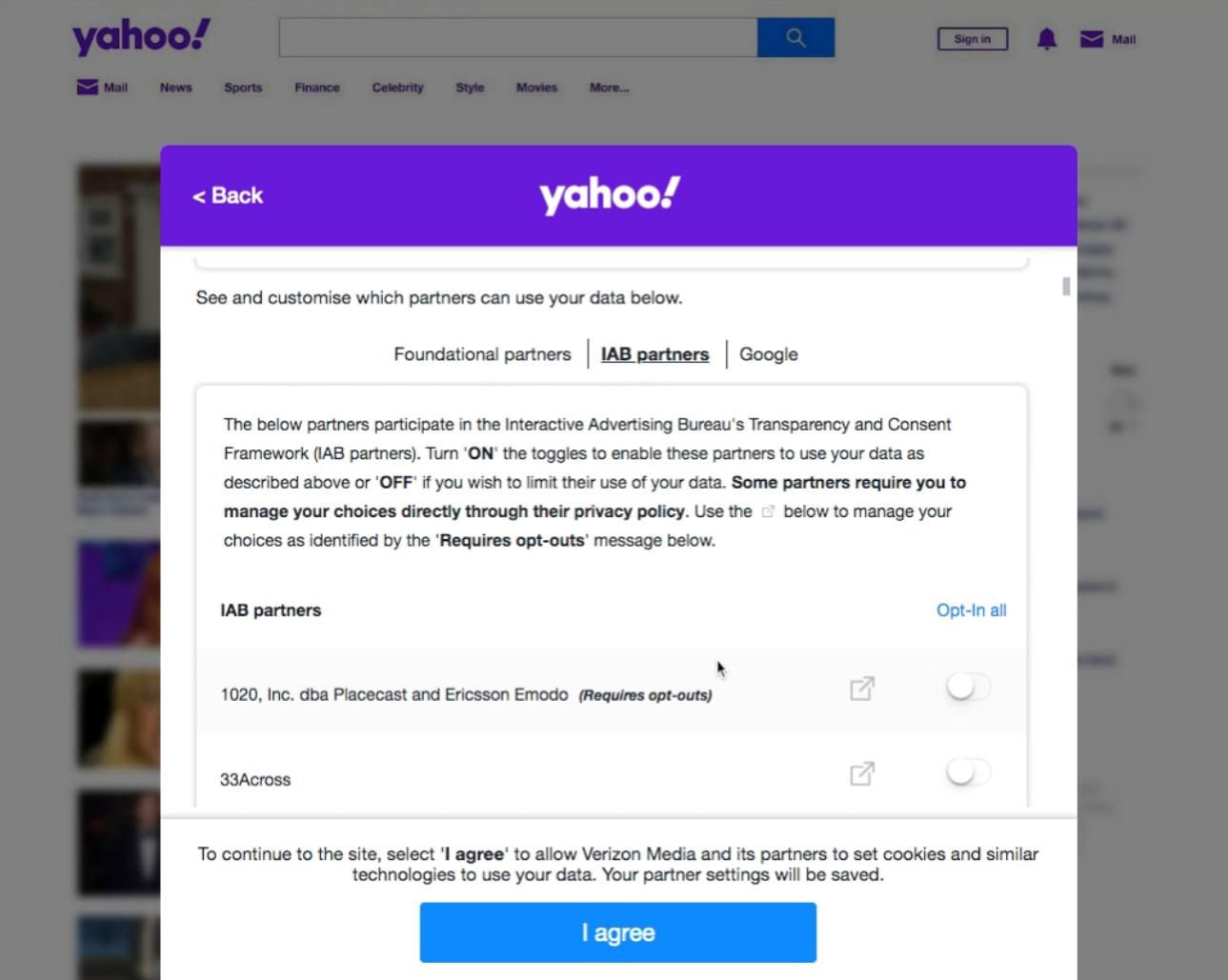}
  \caption{Configuration}
  \Description[Screenshot of Yahoo's website with configuration options available ]{Screenshot of Yahoo's website with configuration options available, including sections for "Foundational partners", "IAB partners", and "Google"}
\end{subfigure}\hfill
\caption{Example of a tracking wall, recorded on 4 March 2020. Credits: Yahoo}
\Description[Three screenshots describing the "tracking wall" on Yahoo's website]{Three screenshots describing the "tracking wall" on Yahoo's website}
\label{fig:Trackingwall}
\end{figure}

A tracking-wall is an instance of a \emph{consent wall}, however with more detrimental consequences to the user. In addition to blocking access to the website until the user makes their choice, a tracking wall gives the user only one option: to consent and accept any terms offered by the site, without any possibility to refuse. In the legal domain, a tracking wall is also called a ``cookie-wall'' or ``take it or leave it'' choice~\cite{Borg-etal-17-EDPLR}. Differently from a \emph{consent wall} (section~\ref{sec:findings-consent-wall}), a tracking wall cannot result in a \emph{reduced service} (section~\ref{sec:findings-reduced}) because the only option the user has is merely to accept consent in order to access the website.
An example of this design choice can be found on the website of \url{https://yahoo.com} which illustrates the use of a consent banner that provides only one choice---to accept---while blocking access to the website, as depicted in Figure~\ref{fig:Trackingwall}\footnote{Video recorded on 4 March 2020: \url{https://mybox.inria.fr/f/6d9ea3b16c6b487d8065/}.}. Our interaction criticism analysis of consent walls provided in section~\ref{sec:findings-consent-wall} applies to the tracking wall as well. In this section, we complement the consent wall analysis with additional specificity related to tracking wall design choices.


\paragraph{Designer perspective} When deploying a tracking wall in a website, a designer chooses to restrict a visitor's access to content or service when that visitor denies consent. 
Therefore, the only access possibility is complete acceptance of tracking technologies used by the website provider and/or their third-party partners, under any terms that may be provided explicitly, hidden, or simply left unstated. As a result, this design choice puts more aggressive pressure on the user's action, with even less respect of a freely given choice. It therefore raises the same questions as the consent wall regarding the legal requirement of a \textit{freely given} consent.
%

\paragraph{Interface perspective} The interface here is very similar to the one encountered when facing a consent wall. The only difference is the absence of a possibility to refuse consent. As shown in the example provided in Figure~\ref{fig:Trackingwall}, some tracking wall examples do include a second informational button (``learn more''); however, even if present, these buttons typically do not provide immediate access to additional configuration options. The overall impact of this interface experience serves to obstruct access to any web resources except for the consent box, until the only real choice of ``I agree'' has been made. 

\paragraph{User perspective} A tracking wall represents a form of obstruction which prevents the user from achieving their intended action, such as reading an article, creating an account, logging in, or posting content. It interrupts the user browsing, giving them a single ``choice'' to give consent or to quit the website. The absence of any way of using a service/accessing a website without giving consent (e.g. via a ``Refuse'' or ``Decline'' option) makes the interface actively coercive, leading to an unpleasant experience for users who do not wish to give consent. Thus, beyond being obstructive, this lack of freely given consent may also constitute a form of forced action. From a legal perspective, the CNIL’s Draft Recommendation on the use of cookies~\cite{CNIL-draft-rec-cookies-2020} proposes that consenting to trackers should be as easy as refusing them, and users should not be exposed to negative consequences should they decide to refuse consent to tracking. 


\paragraph{Social impact perspective} A tracking wall, from a website owner's point of view, could be a means to offset costs relating to providing the web service, facilitating a balancing of traffic with advertising revenues. Choosing to completely block the site has a greater impact than a consent wall, as it is likely to deprive part of the population of access to all the content or service. More specifically, this restriction may make privacy concerns incompatible with the use of a website not financed by the user, such as those financed by advertising. For instance, on information and news websites, this type of design choice may restrict access to information for users depending on their income. In the worst case, this could lead to significant disparities in accessing information and equality among individuals, with the wealthiest people falling back on paid sites without advertising. Paywalls do exist in some areas, they are generally reserved for content where there is a general social understanding of cost.

The majority of the stakeholders and regulators concur that failure to consent to the use of trackers should not result in the restriction of access to the website’s content. However, the legal prohibition of this practice varies by source, with the European Data Protection Supervisor (EDPS)~\cite{EDPS-ePriv}, the European Parliament~\cite{Parliament-proposal-2019}, the Bureau Européen des Unions de Consommateurs (BEUC)~\cite{BEUC-ePriv}, the Dutch~{\cite{Dutch-DPA-cookiewall,Dutch-ministry-cookiewall}}, Belgian~\cite{belgian-dpa-cookies}, German~\cite{german-guidance2019}, Danish~\cite{danish-dpa-consent} the Greek and Spanish DPAs~\cite{spanish-dpa-cookies} all agreeing that this practice should be deemed unlawful. In contrast, the ICO~\cite{ICO-Guid-19} and the Austrian~\cite{austrian-dpa-consent} DPAs diverge on their opinion of the admissibility of tracking walls. In May 2020, the European Data Protection Board (EDPB) addressed the legitimacy of cookie walls and considered~\cite[3.1.2. (39 -- 41)]{WPconsent2020} that the requirement for free consent implies that ``access to services and functionalities must not be made conditional on the consent of a user to the storing of information, or gaining of access to information already stored in the terminal equipment of a user (so called cookie walls)''. Thus, using ``a script that will block content from being visible except for a request to accept cookies and the information about which cookies are being set and for what purposes data will be processed'', with ``[\ldots] no possibility to access the content without clicking on the `Accept cookies' button'' is regarded as non compliant to the GDPR.


\paragraph{Summary} 
Consistent with our analysis of the consent wall, this design choice increases the tension between interactive separation of user activities and the requirement to allow the user to freely give their consent. In addition to this primary design and legal tension, the lack of an ability to reject consent---alongside the inability to use the web resource without making this forced choice---represents an additional barrier to the user's ability to make a specific and informed decision.



\subsection{Configuration and Acceptance}
\label{sec:findings-configuration}

\subsubsection{Reduced Service}
\label{sec:findings-reduced}

\begin{figure}[!htb]
\centering
\begin{subfigure}[t]{0.45\textwidth}
  \includegraphics[width=\linewidth]{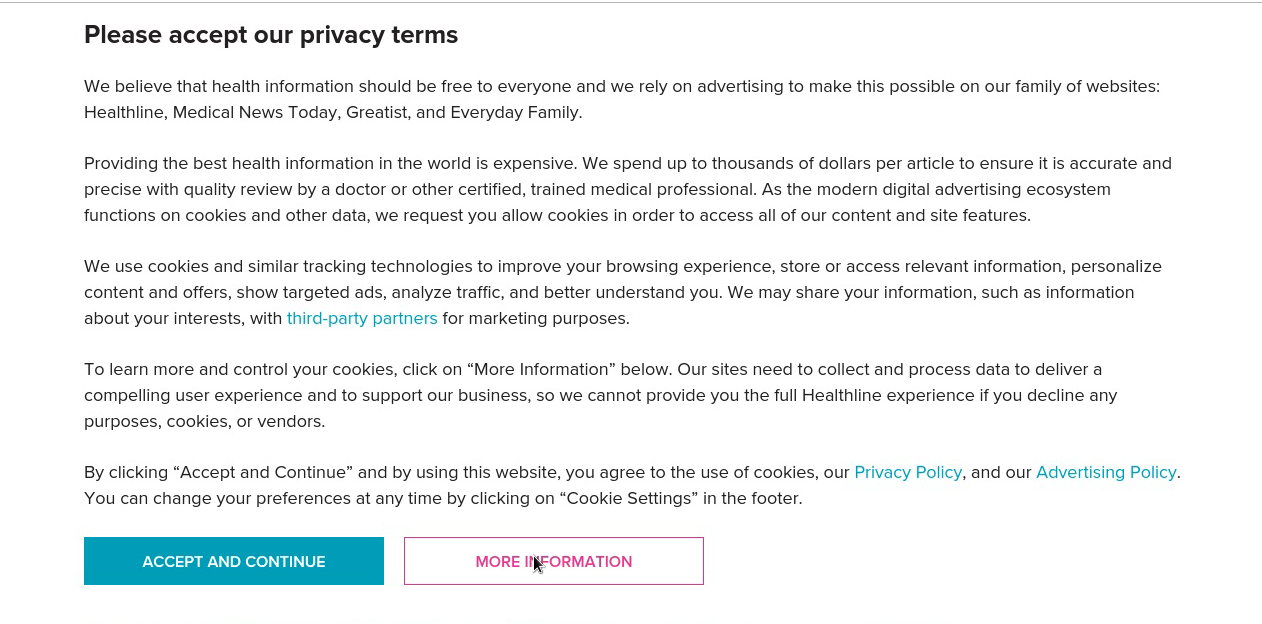}
  \caption{Initial framing}
  \label{fig:ReducedService_1}
  \Description[Screenshot of the initial framing of Healthline's website]{Screenshot of the initial framing of Healthline's website, with only the choices "Accept and continue" or "More information" available to the user}
\end{subfigure}\hfill
\begin{subfigure}[t]{0.45\textwidth}
  \includegraphics[width=\linewidth]{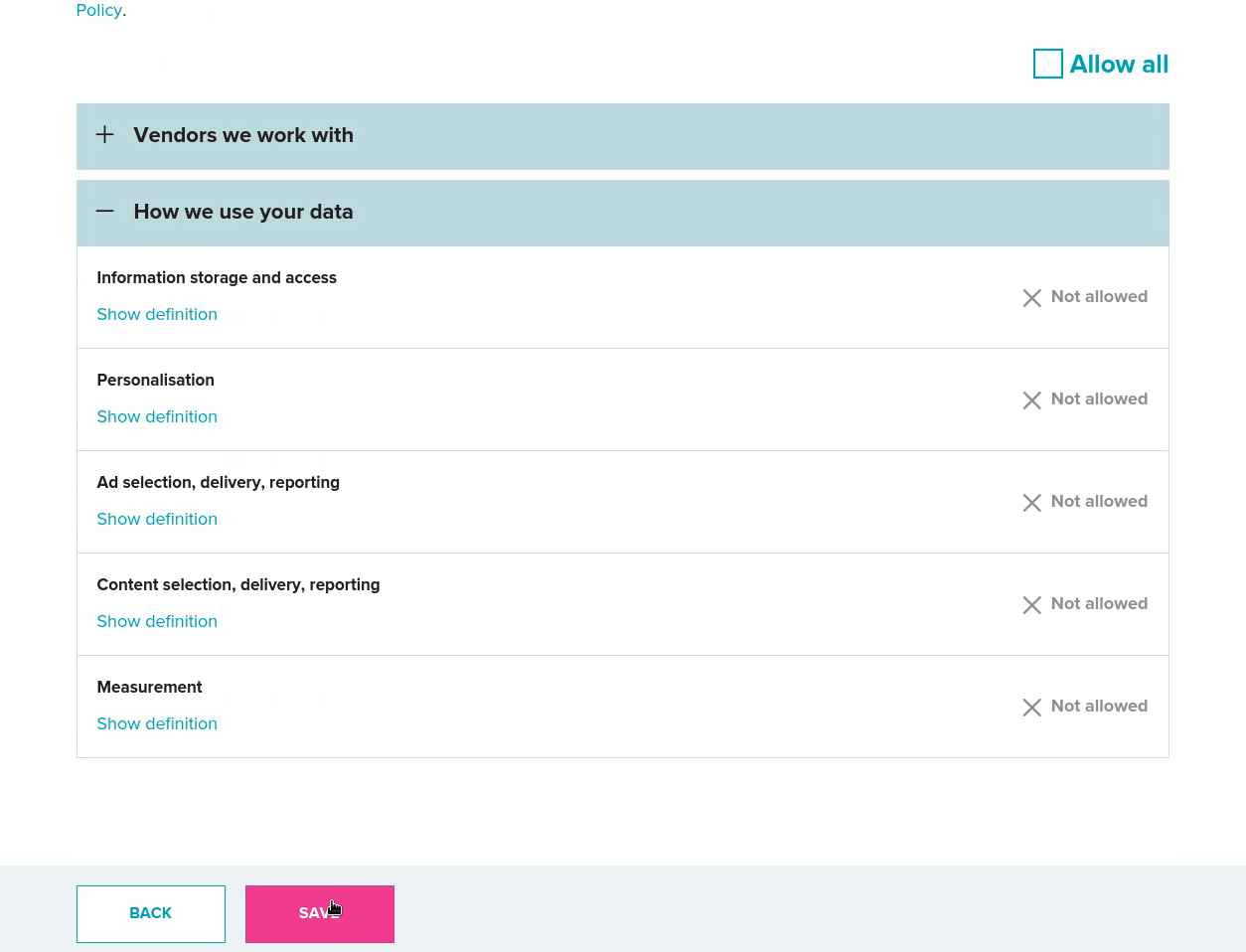}
  \caption{Configuration}
  \label{fig:ReducedService_2}
  \Description[Screenshot of the configuration page of Healthline's consent form]{Screenshot of the configuration page of Healthline's consent form, with multiple choices under the section "How we use your data."}
\end{subfigure}\hfill
\begin{subfigure}[t]{0.45\textwidth}
  \includegraphics[width=\linewidth]{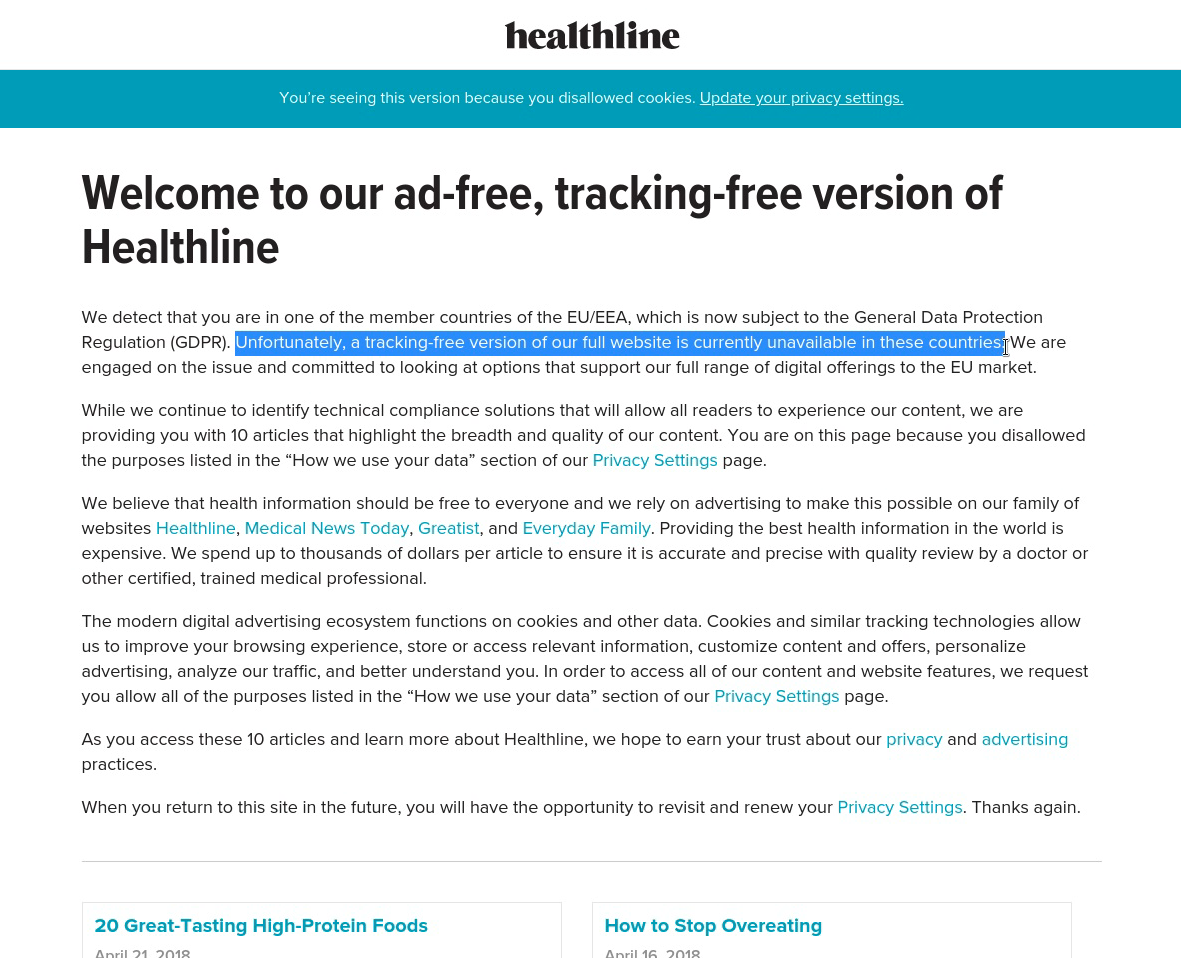}
  \caption{Alternate ``reduced'' version of the website}
  \label{fig:ReducedService_3}
  \Description[Screenshot of the "reduced" version of the Healthline Website]{Screenshot of the "reduced" version of the Healthline Website, which has a headline indicating it is the "ad-free, tracking free version of Healthline."}
\end{subfigure}\hfill
\caption{Example of a reduced service, recorded on 1 October 2019. Credits: healthline.com}
\Description[Three screenshots describing "reduced service" on the Healthline website]{Three screenshots describing the "tracking wall" on Yahoo's website}
\label{reducedservice}
\end{figure}

The use of \textit{reduced service} refers to the practice of a website offering reduced functionality---for example, allowing a user access to only limited number of pages on a website---based on their consent configuration options. In the scope of this paper, reduced service is a result of the user refusing consent in some or all of the proposed privacy configurations. An extreme case of a reduced service occurs when a website fully blocks access because the user refuses some of the privacy configurations.
%
In one example of this design choice, the website \url{https://www.medicalnewstoday.com/} shows that when the user refuses consent, they are redirected to 
another website \url{https://anon.healthline.com/}, which is a reduced version of the original website with only 10 pre-selected pages available to the user, as depicted in Figure~\ref{reducedservice}\footnote{Video \url{https://mybox.inria.fr/f/1ec82ce1a4dd4f82b556/} recorded on 4 March 2020.}. 
Interestingly, if the user visits \url{https://www.medicalnewstoday.com/} again after making this configuration choice, the full website is available. 

\paragraph{Designer perspective.} 
A reduced service consists of a second version of the website with less functionality, differentiating the content made accessible to the users depending on their acceptance or refusal of all or some of the privacy configuration options.
This design decision may reflect the business realities that some designers are forced to consider in relation to stakeholders, such as the role of ad-generated content or other types of tracking that may make publishing certain kinds of content untenable \textit{without} erecting the equivalent of an ineffective paywall. From a \emph{legal perspective}, a reduced service option could be allowed if it clearly enables the user to choose between various options of access. For publishers that provide more means of access (such as free and paid), a reduced service option could be allowed if it clearly lets the user to choose between various options of access. As a boundary condition, the Article 29WP~\cite{WP259} states that refusal of consent must be “without detriment” or “lowering service levels”, though such delineation comes without decomposing what this means in concrete settings, particularly in the digital world. 

%

\paragraph{User and interface perspectives.} The user, when refusing tracking, is redirected to a different, reduced version of the website, and perhaps without knowledge that they made a choice that impacted the content they received. The ultimate effect here is a degraded experience for these users, a practice which the NCC~\cite{NCC-Deceived-18} names \emph{``Reward and punishment''}, explaining that service providers use incentives to reward a correct choice (e.g., extra functionality or a better service), and punish choices they deem undesirable, which in our case entails a refusal of tracking. Thus, the overall effect on the side of the user is either experienced as the dark pattern ``forced action'' or ``obstruction'' if the feedforward action upon selection of a consent option clearly results in the user being directed to a site with reduced service, and ``interface interference'' or ``sneaking'' if the consent interface does not provide adequate feedforward instructions, or otherwise misrepresents the nature of choice in relation to its impact on the user experience. In this design choice, the specific nature of the interface elements are less important than the destination to which the user is sent, and the extent to which the user interface provides guidance to allow the user to make an informed and freely given choice regarding whether they wish to access a full or reduced version of the site. However, a \emph{freely given consent} implies that the data subject could refuse consent without detriment which could be construed as facing significant negative consequences (Recital 42 of the GDPR). 

Moreover, the legal requirement of \emph{informed consent} could be violated under the reduced service design choice.
As argued by the General Advocate Szpunar~\cite{Adv-gen-Szpunar-2020}, a data subject must be informed of all circumstances surrounding the data processing and its \emph{consequences}: \emph{``crucially, he or she must be informed of the consequences of refusing consent''}, including a reduced service. He proceeds by asserting that \emph{ ``a customer does not choose in an informed manner if he or she is not aware of the consequences,''} thus potentially rendering instances where feedforward in the interface is missing to be legally problematic. Additionally, this limitation of service, conditional on consent, obliges the user to give consent to the data processing in order to fully access the website, and therefore, in the absence of another access option, may also violate a freely given consent requirement. In a similar line of thought, Acquisti et al.~\cite{Acqu-etal-17-ACMCS} propose that increasing the cost or the difficulty of choosing specific configurations, even at the simple level of requiring multiple confirmations, configures a ``punishment'' that could prevent inexperienced users from selecting risky settings. 

%

\paragraph{Social impact perspective.}
From a social impact perspective, we start our analysis by considering the intentions of a website owner, pointing towards broader issues of economic viability. Reduced service, from one perspective, could be a response to the economic need of the website owner to find a working business model, thus allowing users to access the full version of the website only if they gave a positive consent, and hence the website can be funded indirectly via data collected from the user. Beyond the technical complexity of presenting two or more versions of the same web property, there are also potential issues relating to archival access of content, deep linking, or other forms of user discovery that have become typical in most web experiences.  Notions of free and unencumbered access is increasingly problematic on the internet, evidenced by resistance to paying for quality journalism and expectations of access to content through bundling with a larger service (e.g., Netflix, Amazon Prime). 

This design choice also points towards potentially relevant legal obligations which are often hidden to end users. The website owner must find a balance between the economic and legal requirements, but the main tool by which they might make this separation may prove to be overly coercive, violating the assumption that consent is ``freely given.'' One way to approach this difficult balance may be to propose users pay for access to the website if consent is refused. However, such paid models lead to further social consequences. A choice between a paid option without tracking and a ``free'' option, financed by tracking, implies that the user's right to privacy is conditioned to paying a fee, which introduces unequal access to a fundamental right to privacy for different categories of users. This raises the question of the compatibility between (1) the obligation to respect users' rights, equality of rights even when users don't have the same level of income, and (2) the need for funding for the website.

\paragraph{Summary.} This design choice presents tensions among separation of access to content based on 1) the consent choice of the user, 2) the economic realities of producing and providing access to content, 3) requirements for consent to be freely given with outcomes that are transparent to the user, and 4) increasing social expectations that web content be accessible without cost or obligation. All these tensions point toward potential acceptance of this design choice, but only in cases where the feedforward interaction---explicitly indicating that certain consent decisions will result in reduced service---is transparent and non-coercive, without the use of sneaking or interface interference dark patterns. However, most instances of this design choice are likely to fail, either by limiting consent choices up front, or by using manipulative language to lull the user into accepting a choice with different consequences than they expect.



\subsubsection{Other Configuration Barriers}\
\label{sec:findings-barriers}

\begin{figure}[htb]
\centering
\begin{minipage}[t]{0.47\textwidth}
\includegraphics[width=\linewidth]{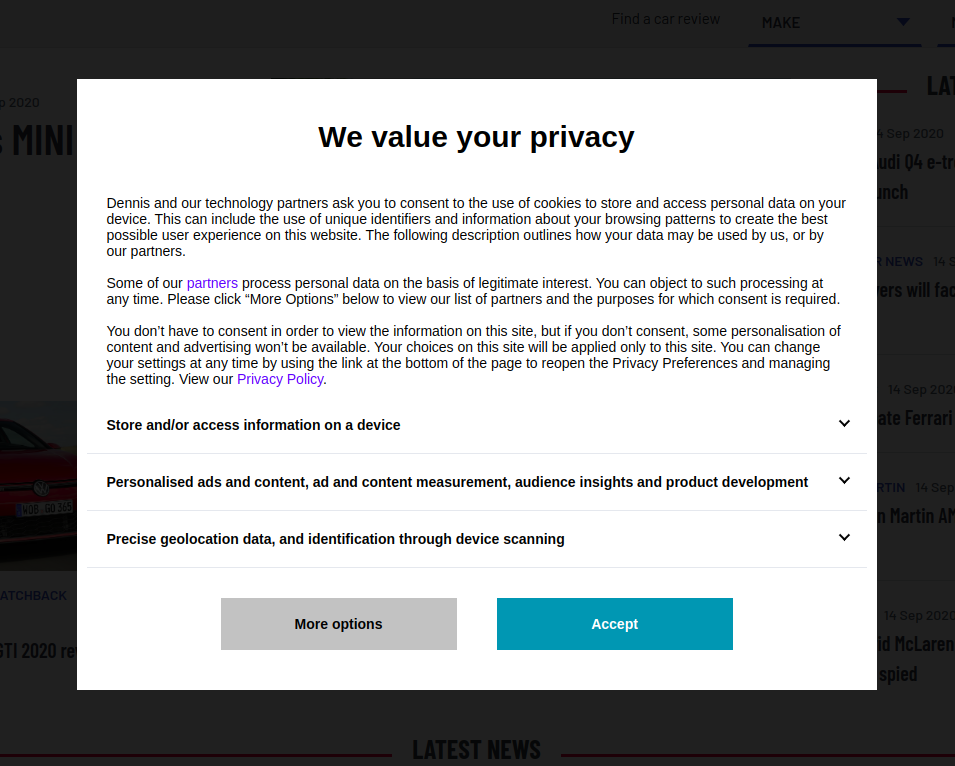}
  \caption{Use of aesthetic manipulation in the
  presentation of consent options (Credits: autoexpress.co.uk).}
  \label{fig:ConfigBarrier_1_Autoexpress}
  \Description[Screenshot of the consent options on Autoexpress.co.uk]{Screenshot of the consent screen on Autoexpress.co.uk, with multiple arrows to other hidden consent options alongside the buttons "More options" and "Accept."}
\end{minipage}\hfill
\begin{minipage}[t]{0.47\textwidth}
  \includegraphics[width=\linewidth]{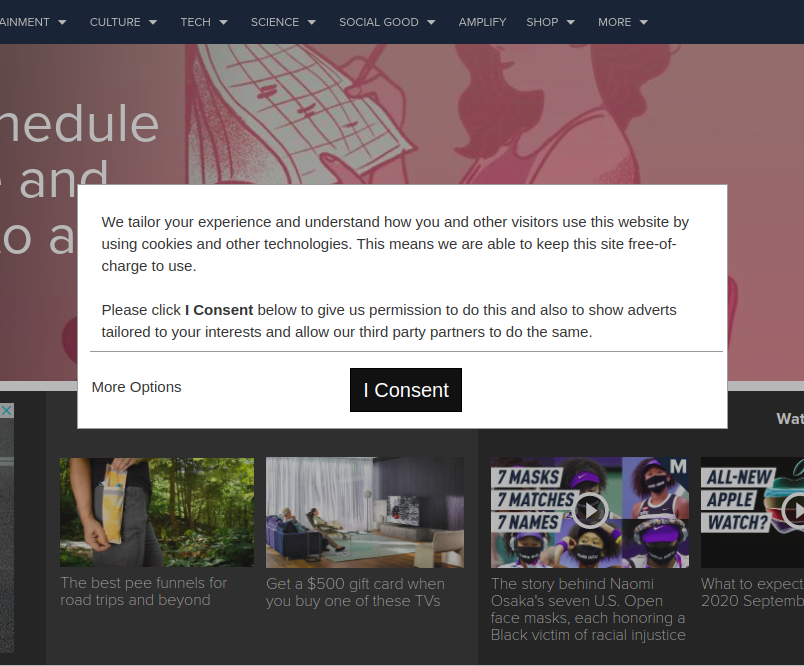}
  \caption{Use of reading order manipulation to discourage certain consent options (Credits: mashable.com).}
  \label{fig:ConfigBarrier_2_Mashable}
  \Description[Screenshot of the consent screen on Mashable.com]{Screenshot of the consent screen on Mashable.com, with "More Options" listed in plain text on the far left of the popup, out of the normal reading order of the main button "I Consent."}
\end{minipage}\hfill
\label{fig:ConfigBarriers}
\end{figure}

Configuration barriers usually correspond to known implementations of consent mechanisms that dynamically interact with the user and direct them towards acceptance of consent \cite{Nouwens2020-ij,Soe2020-se}. Configuration choices can be deconstructed into a variety of more basic design choices, such as: 
\begin{itemize}
    \item The imposition of hierarchies or prioritization of choices which should have instead equal value or positioning. We observe this practice in consent dialogs with a larger ``OK'' button that appears first, and a smaller ``Configure'' button gives a more prominent visual hierarchy to ``OK.''
    \item The introduction of aesthetic manipulation (also known as ``attractors'' or ``interface interference''), where desired and concrete user choices are perceived more salient and prioritized. An example of this phenomenon might include a bright and attractive ``accept'' button and either a gray ``reject'' or ``more options'' button (Figure~\ref{fig:ConfigBarrier_1_Autoexpress}).
    \item The use of reading order manipulation to ``sneak'' information past the user. One example of this includes the use of a box ``I consent'' emphasized in a black box, and ``More Options'' link on the (left) corner of the banner
    , outside of the normal reading order (Figure~\ref{fig:ConfigBarrier_2_Mashable}).
    \item The use of hidden information that is hidden behind another interactive element or otherwise invisible to the user without further investigation. For instance, the use of plain unformatted text to indicate a link to ``Preferences,'' while ``Accept'' is a visible button.
\end{itemize}


Below we consider this range of interrelated design choices as a complex set of visual and interactive design criteria (e.g., \cite{Benyon2014-qx,Cooper2014-ot,Tidwell2010-tq}) from which designers can draw in creating design outcomes that allow users to select consent options.

\paragraph{Designer perspective.} 
The designers intent to use influencing factors that are visually salient (e.g., a larger button for ```accept,'' the use of bright colors for objects with a higher priority, or the use of hovering properties to disguise feedforward) have a clear and direct effect of prioritizing the choice of acceptance of tracking over rejection, even if such choices are less privacy-friendly to the end user. While many of these visual techniques are well known, building upon gestalt psychology principles, and are often used to create more efficient and engaging user experiences, these same principles can be co-opted through the use of ``interface interference''-oriented dark patterns.

\paragraph{Interface perspective} 
Specific configuration properties of cookie banners have been manipulated in order to influence users’ decision to give consent~\cite{Utz-etal-19-CCS}. The means of manipulation include many aspects of visual and interactive display, including the positioning, size, number of choices given, formatting (use of fonts and colors emphasising consent options, widget inequality), hiding settings behind difficult to see links, preselected boxes, and unlabeled sliders. These \emph{attractors} are interface elements that are intentionally designed to draw or force the attention to a salient portion of a larger interactive experience~\cite{BravoLillo2014HarderTI}. The ``salient field'' is the part of the consent dialog that provides the most important information to aid the user's decision. The use of such eye-catching techniques makes it easier to see and act on some design elements than others, and making some buttons or options more salient is an example of design outcomes that are intended to surreptitiously nudge users by making a pre-chosen and intended choice more salient\cite[pp. 19-20]{NCC-Deceived-18}. 
From a \emph{legal perspective,} Article 7(4) of the GDPR states that withdrawing consent should be as easy as giving it, and we additionally interpret that the choice between ``accept'' and ``reject'' tracking must be consequently balanced and equitable and as such, design choices related to an unbalanced choice violate the legal requirement of an \emph{ambiguous consent}.
In fact, 
{``[a] consent mechanism that emphasizes ‘agree’ or ‘allow’ over ‘reject’ or ‘block’ represents a non-compliant approach, as the online service is influencing users towards the ‘accept’ option.'' },~\cite{ICO-Guid-19}. 
The Advocate General of the Court of Justice of the EU ~\cite{Adv-gen-Szpunar-2019} emphasized the need for both actions, ``optically in particular, [to] be presented on an equal footing.'' Thus, while  \emph{the procedure to choose should be as simple as to accept} is legally warranted, pointing towards a series of design choices that makes the acceptance and refusal buttons visually balanced (or equitable), the complex array of design choices in play make the practical inclusion or exclusion of certain interface choices difficult to precisely objectify.


\paragraph{User perspective.}
The apparent need for attractors stems from the fact that \emph{attention is a limited resource}; consumers are often multi-tasking and focusing on many different stimuli at once~\cite{Adjerid2013SleightsOP}. The attentiveness of consumers to privacy issues may be sporadic and limited, inhibiting the usefulness or impact of even simple and clear privacy notices. Therefore the salience of stimuli can impact the user's decision-making processes and outcomes. The configuration practices of \emph{``attention diversion''}~\cite{CNIL-Shap-19} draw attention to a point of the website with the intention to distract and/or divert the user from other points that could be useful. 
The French Data Protection Authority adds that designers can take advantage of user psychology, for instance deciding to make the color of a ``continue'' button green while leaving the ``find out more'' or ``configure'' button smaller or grey. If users are conditioned by the traffic light metaphor bias used by designers that assign colors according to the flow of information (``green'' = free flowing; ``red'' = stop), users may perceive green as the \emph{preferable choice}.

\paragraph{Social impact perspective.}
Services offer a carefully designed interface, which rather than configuring a neutral conduit, instead nudge the user into acting in the best interest of the shareholder. While these behavioral techniques are well known in industry settings, most users are not aware of the degree to which their everyday patterns of use are predetermined, based on knowledge of human psychology in general and the actions of users in particular contexts. Many of the visual and interactive choices indicated above are not neutral, but rather---in combination---have been shown through A/B testing or use of other evaluation to produce the desired output behavior from users. Thus, while societal norms at large might dictate that interfaces should not use potentially misleading design practices---such as the use of visual grammar that might lead the user to think that consent is required to continue browsing, or that visually emphasizes the possibility of accepting rather than refusing---the capabilities of digital systems to rapidly test and deploy interface combinations that are optimized for certain behaviors act against our broader desire as a society to make informed and deliberate choices about how our data is collected and used.

\paragraph{Summary}
These series of overlapping and cascading design choices provide a central point of focus for the desired and actual experience of the consent process. The notion of configuration is central to the ability of the user to make an unambiguous and specific choice about how their data can be collected and used. However, as shown above, so many of the visual and interactive elements relate and interact in ways that resist the ability of policy to specify allowable and unallowable design choices. While some tactics can be used to provide a better user experience (e.g., use of color to indicate the role of different options and their meaning in relation to feedforward interaction), they can easily be subverted as well. Thus, while the outcomes are clear from a legal perspective, it is virtually impossible to demonstrate in full what design choices are relevant, appropriate, and legal---either separately or in combination.

\section{Discussion}
\label{sec:discussions}
In the previous sections, we have identified different approaches to engagement with consent banners across the user task flow, including: a) altering the initial framing of the consent experience through a consent wall or tracking wall; and b) manipulating the configuration and acceptance parameters through reduced service and other barriers to configuration. Using an interaction criticism approach, we described the complex forms of disciplinary engagement and tensions built into each set of design choices as experienced from four different subject positions, including: the designer, the interface itself, the end user, and the broader social impact. In articulating each consent experience from these multiple points of view, we have sought to bring together design, computer science, and legal perspectives, particularly acknowledging instances where these perspectives foreground tensions in satisfying concerns raised from these disciplinary perspectives. Building on our findings---and the many discussions that supported our investigation of consent banners---we present below a further synthesis of our transdisciplinary dialogue. First, we describe how argumentation can be productively commenced and sustained from both design and legal perspectives. Second, we build upon this mode of argumentation to describe new opportunities for dialogue across legal, ethics, computer science, and HCI perspectives to engage with matters of ethical concern through the lens of dark patterns.

\subsection{Bi-directional Design and Legal Argumentation}
\label{sec:discussion-argumentation}


We have demonstrated the value of approaching a complex issue such as the design and regulation of consent experiences from the perspective of multiple disciplines, revealing through our analysis a range of synergies and disconnects between these perspectives. We argue that although there is a desire for standardization, 
enabling the exercise of a valid choice by end-users, there does not appear to be a fully neutral set of design requirements by which operators can guarantee that all elements of the GDPR can be satisfied. When engaging in a bi-directional means of argumentation between design and legal perspectives, we can identify some of the areas of tension and opportunity---pointing to new possibilities for policy implementation, and better ways of managing legal requirements during the design and development process. By ``bi-directional,'' we refer to the opportunities to  evaluate and interrogate designed experiences using the language of law and policy (legal->design), while also using a user experience or user interface as a means of addressing gaps or opportunities for more precision in existing legal or policy frameworks (design->legal). 

\begin{figure*}[ht]
  \centering
  \includegraphics[width=\linewidth]{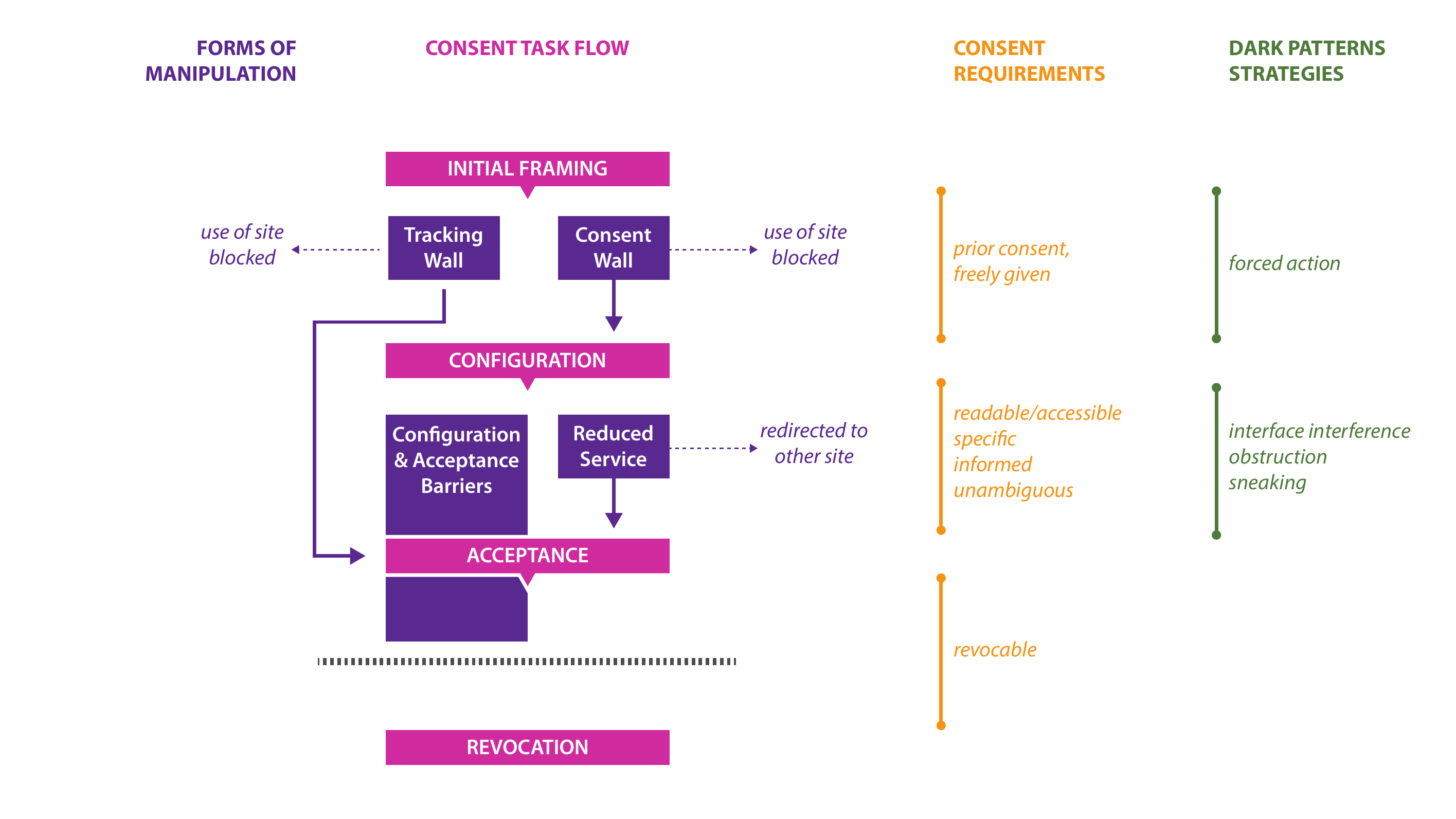}
  \caption{Flowchart describing the forms of manipulation we observed in our dataset in relation to the consent task flow, legal consent requirements, and dark patterns strategies.}
  \Description[Flowchart that aligns the main task flow elements, consent requirements, and dark patterns strategies]{Flowchart with the main task flow elements on the left, consent requirements in the center, and dark patterns strategies on the right. The task flow shows the tracking wall and consent wall under initial framing, resulting in the site being blocked; configuration barriers and reduced service under configuration, with reduced service resulting in redirection to another site. Consent requirements of prior consent and freely given align to initial framing, readable/accessible, specific, informed, and unambiguous aligned to configuration, and revocable aligned with acceptance. Forced action is indicated as a dark patterns strategy for initial framing, while interface interference, obstruction, and sneaking are aligned with configuration.}
  \label{fig:flowchart}
\end{figure*}

Beginning from a legal perspective, we can 
envision the role of standardization in consenting procedures, including a 
list of ambiguous behaviors that must be explicitly acknowledged by decision-makers. 
Ensuring standardization could enable rapid detection of violations at scale (building on similar work in e-commerce by Mathur et al. \cite{Mathur2019-hx}) while also minimizing legal uncertainty and subject appraisal regarding configuration aspects of consent banners. Using a standardization approach could also minimize behaviors presenting a margin of doubt regarding the choice expressed by the user, as advocated in recent work \cite{Cookie-reco-Inria-2020, Santos2019-fe}. When interrogating this desire for standardization from a design perspective, we can see deficiencies in the current data protection framework which do not accurately model or describe relevant HCI, UI, and UX elements when assessing the lawfulness of dark patterns in consent formulation. While previous empirical work acknowledges the impact that HCI and UI provokes in the user's decision-making process (e.g.,~\cite{Grassl2020-lh,Nouwens2020-ij,Soe2020-se}), it is currently unclear whether fully ``neutral'' design patterns exist, and even if they did, how a list of possible misleading design practices that impact both users' perception and interaction that impact a compliant consent could be fully dictated \textit{a priori}.

Using Figure~\ref{fig:flowchart} as a guide, we can begin to identify legal and design/HCI/UX endpoints with which to  
start a conversation, revealing various disciplinary perspectives that have the potential to guide future policy and design decisions. It is clear that even if some of these disciplinary perspectives may appear debatable or blurred when presented through the practice of interaction criticism, the disconnects and synergies signifies instead a space of vitality and opportunity at the nexus of these domains that may point towards patterns that \textit{should} be illegal and patterns that are more \textit{likely} to produce a fully-compliant consent. We propose that developments will be more rapidly identified and consolidated when the transdisciplinary perspectives across legal, ethics and HCI scholarship are integrated within case-law and also validated in academic research. In particular, a reading of this figure allows researchers, designers, and legal scholars to identify spaces where social and political values might successfully emerge together or collide, recognizing that the foregrounding of certain perspectives and language indicates that ethical engagement with these issues is always already political and value-laden. The political and ethical dimensions of design choices requires designers, researchers, and legal scholars alike to use a pragmatist ethics approach to identify and rationalize design choices in relation to legal requirements, user value, and potential or actual societal impact.

\subsection{Opportunities to Bridge Legal, Ethics, and HCI Scholarship}
\label{sec:discussion-bridging}


Building on the need for bi-directional argumentation shown above, we see a opportunity for further interwovenness between design choices and legal guidance, using this liminal space as a means of describing ways to engage in more transdisciplinary ways underneath the conceptual umbrella of ``dark patterns.'' While other conceptual means of connecting these disciplinary perspectives are possible and potentially useful, we will that demonstrate the conceptual unity among these perspectives---bringing together scholars from many disciplinary perspectives--- is possible in the sections delved below.

First, while assessing each consent experience, we observed a \emph{tension} between dark pattern categories because many dark patterns overlap in the different visual and interactive design choices of a single banner~\cite{Soe2020-se}. These patterns are blurred and difficult to distinguish, and in fact, we as human evaluators frequently disagreed on which dark pattern might exist on a specific banner, and from which perspective one evaluation may be more or less tractable. For example, obstruction and interface interference are often perceived at the same time, but are perhaps co-constitutive; interface interference foregrounds a visual design and gestalt psychology perspective, while obstruction foregrounds a view of the temporal user journey and user goals. This difference in perspective reveals that any analysis on design choices and the detection of dark patterns is \emph{interpretive} and therefore there is need for different and combined methods. Though some dark patterns have been successfully detected through computational means \cite{Mathur2019-hx}, many of the aspects of user experience that we highlighted above cannot be easily detected automatically, and may be revealed only through a manual analysis and consideration of multiple user and interactive characteristics. \textbf{This insight reveals that dark patterns is a n-dimensional phenomenon which includes dimensions of time, interaction, design, psychology, and law---demanding a holistic analysis from many perspectives. }

Second, we have revealed that some of the analyzed design choices correspond to known classifications of dark patterns and moreover, they fit neatly within current regulatory structures that prohibit and sanction deceptive practices, as it is the case of tracking walls, which are explicitly forbidden by the European Data Protection Board~\cite{WPconsent2020}. Thus, these design choices should be considered legally actionable and subject to enforcement actions by the competent authorities. Conversely, some design choices might be deemed as unlawful, but fail to fit the threshold of what is mandated by explicit legal requirements, though arguably falling outside of existing data protection regimes (for example, the case of reduced service). 
Regarding the role of design choices that might trigger legal or policy implications, we agree with Schaub et al.~\cite{Schaub-et-al-2015} and Karegar et al.~\cite{Karegar-et-al} which argue that the main problem might not be inherent to the requirements postulated by current legal sources, but in how consent dialogs are currently designed.
This disjuncture in potential outcomes points to two plausible directions for bridging and transdisciplinary discourse: 
a) a pathway towards recognition of design choices that are knowingly applied by designers that are demonstrably causal in producing impact that is negative from a user or legal perspective, and are unnecessarily disruptive in a way that could be deemed unlawful; and 
b) a means of identifying and encouraging discourse among everyday users around practices which are not unlawful \textit{per se}, but which should nevertheless be discouraged. 

In the first case, active empirical work is needed to determine causality in conjunction with identification of other important variables that must be considered to eventually determine the lawfulness of the relevant design choice(s). Therefore, while the current legal regulatory scope regarding dark patterns in these cases might not be sufficient, it could be established through both empirical and practical means. 
The use of stricter regulations for consent banners that prohibit and sanction evidence-based misleading design practices might not be sufficient on their own to reestablish a privacy-friendly environment. Recent experimental work~\cite{Grassl2020-lh} has shown that even after removing a nudging and manipulative design choice, a form of routinised conditioning could still \emph{persist}, ultimately leading users to behave in a certain way, due to an irreflective default behavior, referred as ``effect survival'' by Hertwig and Grune-Yanoff~\cite{Hertwig2017NudgingAB}. Notwithstanding, the incoming ePrivacy Regulation~\cite{Parliament-proposal-2019} might install a ``Do not Track'' mechanism that would be mandatory for all sites, limiting the number of times users are asked to consent to tracking.

In the second case, broader public and professional participation may be needed to identify negative practices, facilitating users to ``name and shame'' companies that use these patterns and professionals to identify such patterns as irresponsible or destructive within codes of ethics or other constraining professional criteria, as originally proposed by Brignull in relation to dark patterns \cite{Brignull2015-il}. 
Such developments reflect the point that designers are increasingly required to respond with ethically-valenced decisions beyond what may be strictly provided for within legal frameworks and that these design decisions are not neutral, but rather reveal the assignment of value and power. \textbf{More transdisciplinary collaborative research and engagement is needed to translate such abstract debates into practical policy or professional outcomes and to prevent any potential ``moral overload'' in relation to the difficult decisions requiring complicated trade-offs and reflection.}

Third, we have shown that illegal and unlabeled dark patterns can emerge from new analysis, building on the work of Soe et al.~\cite{Soe2020-se} and Matte et al~\cite{Matt-etal-19-SP} and our own application of interaction criticism. For example, the design choices ``consent wall'' and ``reduced service''---while relying upon the dark patterns of obstruction and forced action---are not included in pre-existing categorizations of dark patterns, as defined by others \cite{Bosch2016-vc,Brignull2013-iu,Gray2018-or,Mathur2019-hx, Utz-etal-19-CCS, CNIL-Shap-19}, but rather they emerged from a discussion between legal experts, designers, and computer scientists who are the authors of this paper.
We find it likely that there may be many other types of dark patterns that can be revealed when users interact with consent banners, along with many other means of engaging with data privacy and security. In contrast to this discovery of the ``darkness'' of user interactions, we also present the opportunity to identify new ways to empower users through ``bright'' or ``light'' patterns~\cite{Hertwig2017NudgingAB}, even though empirical research has rendered such pro-privacy nudging approaches as implausible for companies to implement since they are incentivized by tracking user's online behavior. 
One path towards patterns that result in empowerment, supporting the notion of data protection by default and by design (Article 25 of the GDPR), could be accomplished by making the user's decision to share personal information more meaningful---a technique that Stark~\cite{Stark2014-iu} refers to as ```data visceralization'---making the tie between our feelings and our data visible, tangible, and emotionally appreciable.'' In this latter case, \textbf{we point towards the potential role for HCI, UI, and UX designers to work in concert with computer scientists and data privacy experts to further reflect the needs of users into technology design to respond to the regulatory challenges in a more contextually aware manner. }

Fourth, we have raised the question of whether the end-user should solely be considered a central to the decision-making process, and if it is a defensible choice to create this burden and expect a reasoned and fully-informed choice \textit{only} from the user. We posit that the GDPR places substantial---and perhaps unwarranted---pressure on the user by defining the act of consent as a legal basis for processing personal data via tracking technologies.
The definition of consent itself places the burden of choice on the user (through unambiguously given consent) and therefore pressure on the user as well. Such weight comes in the form of a design of the consent interface that a user faces when browsing the internet on a daily basis, and in the long term ramifications of the consent choice, which are never fully knowable. Such an assessment happens in often complex decision-making \emph{contexts} where information is processed quickly, choices abound, and cognitive effort is demanded for the user, making this space a prime opportunity for companies to include dark patterns to encourage certain choices and discourage others~\cite{Stigler-report-2019}. Some user-centered approaches to ameliorate the problems found in the current consent system have been studied, such as the use of ``bright patterns'' and ``educative nudges'' in combination~\cite{Grassl2020-lh}.  \emph{``Bright patterns''} (also known as ``non-educative nudges'') have been used to successfully nudge users towards privacy-friendly options, but these approaches lead to similar problems as their dark counterparts, namely an unreflective default behaviour and users’ general perception of a lack of control. The use of \emph{``educative nudges''} could also be used as reminders or warnings, providing feedback about possible consequences of a user's choice when consenting, however, as the majority of the companies have incentives to track users---nudging them through privacy-unfriendly options---the practical feasibility of such nudges is questionable. Given our experience in working in the legal, computer science, and design fields, we have observed how design choice architectures relying on dark patterns can influence user consent agreements on the data collection and usage in web tracking and that such design choices raise important legal consequences.  We raise the question if potentially there are other ways to make a choice that does not rely on solely on consent as it is currently understood, but on another legal basis, deviating some or all the attention from the end-user.
\textbf{The deeper we look at consent mechanisms and the matter of user choice, the more we understand the need to combine the perspectives of different fields (e.g., HCI, design, UX, psychology, law) as part of a transdisciplinary dialogue in order to ensure that the user's choice indeed satisfies all the consent requirements to be deemed valid: free, informed, specific, unambiguous, readable and accessible. 
}

\section{Implications and Future Work}
\label{sec:future-work}

This analysis points towards multiple productive areas for further investigation of the intersections and synergies of legal, ethics, and HCI perspectives on privacy. 
First, we propose new connections among policymakers and HCI scholarship, building on the work of Spaa et al. \cite{Spaa2019-vm} in identifying ways ``to harness the more speculative and co-productive modes of knowledge generation that are been innovated on by HCI researchers to become part of governmental policymaking processes.'' This effort could be supported by attending in more detail to the ways in which ethical concerns are languaged, with new scholarship mapping opportunities to connect design concepts, notions of design intent, and opportunities for policy to be crafted. Second, the interaction criticism approach we have taken in this paper highlights the value of thinking and interacting with design artifacts across multiple disciplinary perspectives, including transdisciplinary means of thinking through, verbalizing, and conceptualizing design evidence and argumentation. This means of criticism connects with broader goals for design and HCI education and research, including the need for individuals in a transdiscipline such as HCI to be able to raise, respond to, and encourage discourse around multiple disciplinary perspectives. While we cannot claim our application of interaction criticism relating to consent concerns as a distinct methodology for transdisciplinary research engagement from this study only, but we have identified specific aspects of disciplinarity and conceptual vocabulary through the use of interaction criticism---both in terms of productive tensions and means of working out aspects of complexity---that have proven to be useful in building a shared language among our varying disciplinary backgrounds that may be helpful in supporting future transdisciplinary work. More research that focuses on the pathways to building competence in this transdisciplinary dialogue---including the ability to raise both synergies among disciplinary perspectives, and also identify disconnects between language, outcomes, and means of argumentation---could productively reveal best practices for educating the next generation of HCI and UX designers and researchers. Third, perhaps the strongest space for further work is in the integration of legal argumentation in design work, as a means of guiding design practices and as a way of extending and productively complicating legal and policy work. The use of speculative modes of argumentation and interrogation of design artifacts, as proposed by Spaa et al. \cite{Spaa2019-vm}, could lead to the creation of better policies that account for potential futures rather than only deterring known practices. This is an opportunity both to extend the purview of design work, as well as a way of better connecting epistemologies of design and law together in ways that lead to positive societal impact. 

\section{Conclusion}
\label{sec:conclusion}

In this paper, we present an analysis of consent banners through the \textit{interaction criticism} approach, with the goal of bringing together the language and conceptual landscape of HCI, design, privacy and data protection, and legal research communities. Through our analysis, we have demonstrated the potential for synergies and barriers among these perspectives that complicate the act of designing consent banners. Using the language of dark patterns, we have shown the potential for argumentation across legal and design perspectives that point towards the limitations of policy and the need to engage more fully with multiple perspectives of argumentation. Building on our analysis, we identify new ways in which HCI, design, and legal scholarship and discourse may be productively combined with the goal of translating matters of ethical concern into durable and effective public policy.

\begin{acks}
This work is funded in part by the National Science Foundation under Grant No. 1909714, ANR JCJC project PrivaWeb (ANR-18-CE39-0008), and by the Inria DATA4US Exploratory Action project.
\end{acks}

\bibliographystyle{ACM-Reference-Format}
\bibliography{consent}

\end{document}